\documentclass[twocolumn]{emulateapj}
\usepackage{natbib}

\shortauthors{Agol et al.}
\shorttitle{Transits and Secondary Eclipses of HD~189733b}

\begin{document}

\title{The climate of HD 189733b from fourteen transits and eclipses measured by Spitzer} 

\author{Eric Agol\altaffilmark{1,2}, Nicolas B. Cowan\altaffilmark{1}, 
Heather A.\ Knutson\altaffilmark{3}, Drake Deming\altaffilmark{4}, 
Jason H.\ Steffen\altaffilmark{5}, Gregory W.\ Henry\altaffilmark{6},
and David Charbonneau\altaffilmark{7}}

\altaffiltext{1}{Dept.\ of Astronomy, Box 351580, University of Washington, Seattle, WA 98195} 
\altaffiltext{2}{Kavli Institute for Theoretical Physics and Department of
Physics, Kohn Hall, University of California, Santa Barbara, CA
93110}
\altaffiltext{3}{Dept.\ of Astronomy, 601 Campbell Hall, University of California,
Berkeley, CA 94720}
\altaffiltext{4}{NASA's Goddard Space Flight Center, Planetary Systems Laboratory, Code 693, Greenbelt, MD 20771, USA}
\altaffiltext{5}{Fermilab Center for Particle Astrophysics, P.O.\ Box 500 MS 127, Batavia, IL 60510}
\altaffiltext{6}{Center of Excellence in Information Systems, Tennessee State University, 3500 John A. Merritt Blvd., Box 9501, Nashville, TN 37209, USA}
\altaffiltext{7}{Harvard-Smithsonian Center for Astrophysics, 60 Garden St., Cambridge, MA 02138}

\begin{abstract}
We present observations of six transits and six eclipses of the transiting
planet system HD 189733 taken with the {\it Spitzer Space Telescope}
{\it IRAC} camera at 8 microns, as well as a re-analysis of previously
published data.  We use several novel techniques in our data analysis,
the most important of which is a new correction for the detector
``ramp" variation with a double-exponential function which performs
better and is a better physical model for this detector variation.
Our main scientific findings are: (1) an upper limit on the 
variability of the day-side planet flux of 2.7\% (68\% confidence); 
(2) the most precise set of transit times measured for a transiting planet, 
with an average accuracy of 3 seconds; (3) a lack of transit-timing variations, 
excluding the presence of second planets in this system above 20\% of the
mass of Mars in low-order mean-motion resonance at 95\% confidence; (4) a 
confirmation of the planet's phase variation, finding the night side is 
64\% as bright as the day side, as well as an upper limit on the night-side 
variability of 17\% (68\% confidence); 
(5) a better correction for stellar variability at 8 micron causing
the phase function to peak 3.5 hours before secondary eclipse, confirming
that the advection and radiation timescales are comparable at the 8 micron photosphere;
(6) variation in the depth of transit, which possibly implies variations
in the surface brightness of the portion of the star occulted by the planet,
posing a fundamental limit on non-simultaneous multi-wavelength transit 
absorption measurements of planet atmospheres; (7) a measurement of the infrared limb-darkening
of the star, which is in good agreement with stellar atmosphere models;
(8) an offset in the times of secondary eclipse of 69 seconds, which
is mostly accounted for by a 31 second light travel time delay and 33
second delay due to the shift of ingress and egress by the planet hot spot;
this confirms that the phase variation is due to an offset hot spot on
the planet; (9) a retraction of the claimed eccentricity of this system due
to the offset of secondary eclipse, which is now just an upper limit;
and (10) high precision measurements of the parameters of this 
system.  These results were enabled by the exquisite photometric precision 
of the {\it Spitzer IRAC} camera; for repeat observations 
the scatter is less than 0.35 mmag over the 590 day time scale of our
observations after decorrelating with detector parameters.
\end{abstract}

\keywords{stars: planetary systems}

\section{Introduction}

The planet system HD~189733 \citep{Bouchy2005} is one of the best studied 
transiting planet systems due to two factors: its close proximity to our Solar System,
making its star one of the brightest transit host stars, and the large
size of the planet relative to the star, making the transits particularly
deep.  After the secondary eclipse was first detected for this planet 
by \citet{Deming2006}, \citet{Knutson2007} made a precise measurement of the 
phase variation of the planet over slightly more than half of an orbital
period using the 8 micron Infrared Array Camera \citep[IRAC,][]{Fazio2004} on
the {\it Spitzer Space Telescope} \citep{Werner2004}.   In addition to
yielding a longitudinal map of the planet \citep{Cowan2008} which indicated
an offset peak in brightness, attributed to advection of energy by a super-rotating
equatorial jet \citep{Showman2002,Cooper2005}, this observation also yielded the most 
precise measurement of the depth of secondary eclipse, as well the most
precise times of transit and secondary eclipse, for any extrasolar planet.
This motivated us to propose additional observations of six transits
and six eclipses of this system with the goals of looking for secondary
eclipse variability \citep[e.g.][]{Rauscher2007}, looking for transit
timing variations due to other planets in the system \citep{Agol2005,Holman2005},
improving the measurement of the atmospheric absorption \citep[e.g.][]{Tinetti2007},
and improving the measured system parameters for better characterization of 
the planet, host-star, and orbit properties \citep{Winn2007,Torres2008,Pont2007}.

The favorable properties of HD 189733 have allowed detections of planet absorption 
and emission features, yielding possible evidence for water, sodium, methane, 
carbon dioxide, and carbon monoxide, as well as Rayleigh scattering at blue wavelengths 
\citep{Grillmair2007,Tinetti2007,Barnes2007,Redfield2008,Barman2008,
Swain2008,Swain2009,Charbonneau2008,Sing2009,Madhusudhan2009,Pont2008,
LecavelierDesEtangs2008}. 

Despite being an active star \citep{Moutou2007,Henry2008} which affects radial 
velocity measurements, the planet mass is measured precisely to be $M_{\rm p} = 1.13 
\pm 0.03 M_{\rm Jupiter}$, while the radial velocity measurements constrain the 
eccentricity to be $e<0.008$ \citep{Boisse2009}. The planet radius is slightly larger 
than that of Jupiter \citep{Bakos2006a,Baines2007,Winn2007}.  The orbit 
of the planet is well aligned with the spin axis of the star \citep{Winn2006,Triaud2009}.

The deluge of observational constraints on this system has inspired
a wide range of theoretical modeling.  In particular, the
measured phase variation can be qualitatively explained by
general circulation models, such as \citet{Showman2009a} and
\citet{Rauscher2010}, while two-hemisphere models have more
difficulty explaining the shape and variation of the phase
function \citep{Burrows2008}; see the recent review by 
\citet{Showman2009b} for a detailed discussion of these models.

After describing our observations (\S \ref{observations}), we 
give an account of our preliminary data reduction (\S 
\ref{datareduction}), describing our outlier rejection
(\S \ref{outlier}) and choice of centroiding algorithm (\S
\ref{centroid}).   In section 
\ref{photometry} we discuss aperture photometry, background
subtraction (\S \ref{background}), and then detail our
correction for the detector ramp variation (\S \ref{ramp}),
including a new double-exponential model for the ramp (\S
\ref{newramp}) and its performance (\S \ref{rampperformance}).
We complete the description of data reduction with a discussion
of our choice of aperture size (\S \ref{aperturesize}), 
conversion to Barycentric Julian Date (\S \ref{barycentric}),
and error analysis (\S \ref{erroranalysis}).

With the preliminary fit for the ramp, we then simultaneously
fit to the stellar variability and planet variability, outside
of eclipse or transit, and demonstrate the high precision
of {\it Spitzer} IRAC (\S \ref{stellarvariability}).
We then fit the photometry with transit and eclipse models (\S \ref{model}),
and show that the secondary eclipse depth offset can be
explained by light-travel time and the offset hotspot (\S \ref{timingoffset}).
We compute a new ephemeris from our data (\S \ref{timingsection}), and use
the times of transits to place limits on the presence of
companion planets (\S \ref{ttvsection}).
We show that the transit depth appears to vary, which we hypothesize
is due to variations in the stellar surface brightness within the path 
of the planet (\S \ref{transitdepth}) and we show that the day-side planet
flux measured from the secondary eclipses appears not to vary within
the uncertainties (\S \ref{eclipsedepth}).  We discuss these results and
compare to models in the conclusions (\S \ref{conclusions}).

A preliminary analysis of these data were presented in
\citet{Agol2009}; however, we have since made significant improvements
in the analysis, in particular an improved ramp function, so the
results presented here are more reliable.  These data
were also used by \citet{Carter2010} to place a constraint on the
oblateness of HD 189733b, while for the purposes of this paper
we assume the planet to be spherical.

\section{Observations} \label{observations}

We were awarded {\it Spitzer} Guest Observing time during Cycle 4
to observe six transits and six secondary eclipses of HD~189733
with IRAC Channel 4 (PI: E.\ Agol, program ID 40238).  For each
visit we obtained 44,160 exposures of 0.4 second each over 5 hours each.
We also re-analyze the transit and eclipse from \cite{Knutson2007} 
for a total of seven eclipses and seven transits.  We chose IRAC channel 4 
(8~$\mu$m) as it has been demonstrated to be the most stable IRAC band 
\citep[e.g.][and references therein]{Cowan2007}.  Due to the brightness 
of the host star we made the observations in sub-array mode; this
mode allows shorter exposure times (0.4 sec) and faster readout, but
sacrifices the larger field of view (32$\times$32 pixels rather than
256$\times$256 pixels, where each pixel is $1\farcs2$).  We turned off 
dithering which is required for high precision photometry due to the 
array-dependent sensitivity and detector ramp.  We carried out
aperture photometry with a range of radii from 1 to 7 pixels in 1/2
pixel increments.

\section{Data Reduction} \label{datareduction}

We performed our data reductions starting with the Basic Calibrated
Data (BCD) processed with version S16.1.0 of the {\it Spitzer IRAC} pipeline.
These data are corrected for dark current, flat field variations,
and detector non-linearity; they are also converted to units of flux in 
mega-Jansky per steradian (MJy/sr).  After downloading and organizing the data, 
we first converted the images to units of photon counts (i.e.\ electrons) 
by multiplying by the gain (fits header keyword {\sc GAIN} = 3.8 e$^-$/DN) 
and exposure time ({\sc EXPTIME} = 0.32 sec), and dividing by the flux 
conversion factor ({\sc FLUXCONV} = 0.2021 MJy/sr per DN/sec).  For our 
0.32 sec exposure time this amounts to multiplying each pixel by 6.01682 
e$^-$ per MJy/sr.  The elapsed time per exposure
is {\sc FRAMTIME} = 0.4 seconds due to a 0.08 second readout.  We did
not apply corrections for variation in pixel area or corrections to
the flat field for a stellar source; however, we did estimate the
impact of these corrections, and found them to be negligible.

\subsection{Outlier rejection} \label{outlier}

After conversion to counts, we flagged and cleaned the images of 
outliers, such as cosmic rays.  
We cleaned the images at the pixel level by rejecting outliers in the 
time series for each pixel.  This worked 
well since the telescope pointed at nearly the same location 
for the entirety of each of our observations, so the flux in each pixel 
stayed relatively constant making outliers easy to flag.  The pixel flux
is affected by pointing variations, discussed in section \ref{centroid},
so we did the outlier rejection by taking the difference between the 
pixel time series and a 5 exposure (2-second) running median of the time 
series.  The duration of the median was chosen to be shorter than the 
shortest of the pointing excursions, one of which occurred in the middle 
of the transit of the phase-function observation with a 1-pixel pointing
change over 4 seconds.
  
For each pixel, the median-subtracted time series was sorted.  Then, the 
standard deviation was computed from the 68.3\% confidence limits on the 
median-subtracted time series.  Finally, outliers were flagged in the 
median-subtracted time series that differed by more than 4-$\sigma$ from 
zero.  The flagged pixels were then replaced by the 5-exposure median.

This procedure performed well in eliminating cosmic rays
and rogue pixels.  After it was carried out, the photometry showed
no significant outliers and, as we show below, was very close to the 
photon-noise limit.

\subsection{Centroiding} \label{centroid}

{\it Spitzer} is affected by pointing variations that cause the
star to change position slightly on the detector;  this requires
accurate centroiding to perform precise photometry.  There are
four varieties of pointing variations we observe in the data: small 
amplitude, short-timescale ``jitter" which
appears by eye as a damped random-walk; a periodic pointing fluctuation which occurs 
on a time scale of $\sim 1$ hour with an amplitude of about 0.1 pixel; gradual 
drifts over long timescales; and occasional short time scale
sharp excursions, as mentioned in section \ref{outlier}.  

In the inital stages of our data reduction, we found that the
photometry was extremely sensitive to the pointing drifts.
Since the 8 micron camera is undersampled, variations in pointing
at the $\sim 0.1$ pixel level can lead to $\sim$ 10\% variations 
in the flux of the central pixels.  For large photometric apertures, 
this is not a problem since small changes in the centroid do not 
change the enclosed flux much.  However,
large apertures contain pixels with lower illumination which
have a longer-lasting detector ramp (see section \ref{ramp}), so a 
smaller aperture containing higher illumination pixels is more 
desirable since these pixels have a ramp that saturates more 
quickly.  Thus, we realized that a very accurate centroiding 
algorithm would be necessary for using smaller apertures, so we 
set out to test a wide range of different centroiding algorithms 
to see which performed best.  We tried several centroiding algorithms:
flux-weighted centroiding \citep[e.g.][]{Knutson2007}, parabolic 
fitting \citep[e.g.][]{Todorov2010}, Point Response 
Function (PRF) fitting \citep[e.g.][]{Laughlin2009}, and 2-D 
Gaussian-fitting \citep[e.g.][]{Desert2009}.
The simplest algorithms to implement are the first
two since these involve no optimization; the last two involve
iterative non-linear optimization, but the PRF fitting turned out to be
too slow and problematic to implement.

We first tested the centroiding algorithms by creating simulated jitter
using the IRAC channel 4 PRFs. From these
tests we found that the 2-D Gaussian fitting gave the least scatter
in the derived centroid relative to the input centroid; we found
that keeping the $x$ and $y$ standard deviations the same gave
as good a fit to the centroid as allowing the two to vary
independently.  We next ran the algorithms on the two stars 
\citep[target star and M-dwarf companion;][]{Bakos2006b} in the 
phase-function data from \citet{Knutson2007}
and on the two stars in the observations of HD 80606
\citep{Laughlin2009}. These tests were critical since these
were long time series so the stars had time to drift a significant
fraction of a pixel, and the data contain noise.  We first computed 
the centroid of both stars in each image ($x_{\rm 1}$,$y_{\rm 1}$ and $x_{\rm 2}$,$y_{\rm 2}$), 
then subtracted the $x$ and $y$ coordinates for each pair of stars
($x_{\rm 1}-x_{\rm 2}$ and $y_{\rm 1}-y_{\rm 2}$), and finally
binned the $x$ and $y$ differences until the standard deviation of the $x$ and
$y$ differences reached a minimum.  A perfect centroiding algorithm 
ought to have perfect tracking between the two stars, resulting in 
a standard deviation due only to the photon shot noise and
finite spatial resolution of the instrument.  Various choices can be made 
for each of these algorithms, such as what portion of the array to 
fit or whether to smooth the data first before centroiding, so 
we spent some time experimenting with these and other choices.

\begin{figure}[htb]
\centering
\resizebox{8cm}{!}{\includegraphics[width=\hsize]{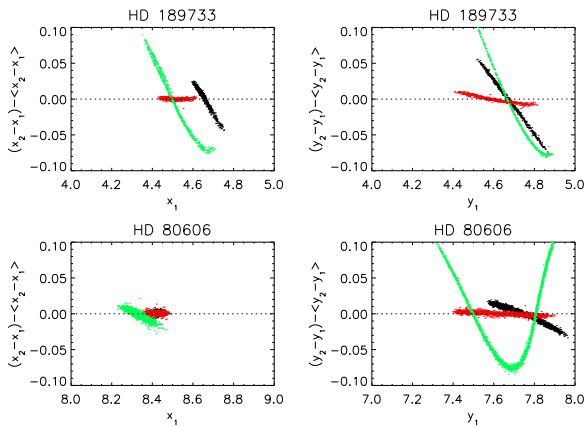}}
\caption[]{Comparison of the centroiding algorithms for HD 189733 versus
M dwarf companion (top panels) and HD 80606 versus HD 80607 (bottom
panels) in the $x$-direction (left panels) and $y$ direction (right
panels).  The black dots are for the flux-weighted centroid, the
red dots for the 2-D Gaussian centroid, and the green dots for
the parabolic centroid.  The HD 189733 data have been binned by
512 exposures (205 seconds), while the HD 80606 data have been
binned by 4 exposures (56 seconds).} \label{fig01}
\end{figure}

In short, we found that the 2-D Gaussian performed the
best of all the centroiding algorithms.  The algorithm selects
a $7\times 7$ sub-array from the image centered on the brightest
pixel of a star.  It then fits a 2-D Gaussian to this sub-array,
allowing the center (centroid), amplitude, and width to vary; four free
parameters in all for each image.  We used
the {\tt mpcurvefit.pro} routine which implements a non-linear
Levenberg-Marquardt algorithm to optimize these parameters 
\citep{Markwardt2009}.  For the HD 189733 phase-function observations,
the scatter in the data were 0.0018 pixels in $x$ and 0.0051
pixels in $y$ when binned by 512 exposures (205 seconds), 
while for HD 80606 the scatter was 0.0015 and
0.0021 in $x$ and $y$ when binned by 4 exposures (56 seconds); 
further binning resulted in minimal decrease
of the scatter.  The second best centroiding
algorithm was the flux-weighted algorithm which had standard
deviations of 0.016 and 0.032 in $x$ and $y$ for HD 189733,
and 0.0019 and 0.011 in $x$ and $y$ for HD 80606.  Thus,
the 2-D Gaussian centroid performed better by a factor of
$\sim 5$ than the flux-weighted centroid; not only that, but
the scatter in the flux-weighted centroid is due to a systematic 
error, while the scatter in the 2-D Gaussian 
centroid is almost completely random.  This can be seen in
Figure \ref{fig01} which shows that the 2-D Gaussian centroid 
has weaker correlation of centroid difference of the two stars versus
the centroid of one of the stars; on the other hand
the other centroiding techniques show signficant correlation
between the offsets of the two stars and the pixel
position of one of the stars, indicating a systematic
error in the centroid determination.

Applying these centroiding algorithms to our twelve transits 
and eclipses and the transit and eclipse from \citet{Knutson2007}, 
we find that the scatter in the difference
in centroids of the two stars (adding in quadrature the
$x$ and $y$ components) ranges from 0.0035 to 0.0042 pixels
for the Gaussian centroid after binning by 128 exposures 
(51 seconds); this is a factor of $3-7$ times smaller than 
the flux-weighted centroid, and also no systematic trend 
in $x$, and a weak systematic trend in $y$.  Since the 
Gaussian centroids of the two stars track 
one another well and the difference in their positions is 
nearly uncorrelated with their position on the detector, we 
are confident that the Gaussian centroid is giving
the correct absolute position of these stars.  When the
data are fit with a 3.5 pixel radius aperture (as described in more
detail below), the 2-D Gaussian centroid yields a $\chi^2$
which is smaller for 9 of 14 transits/eclipses than the
flux-weighted centroid, while the total $\chi^2$ is smaller
by $130$ (after discarding the first 55 minutes of
data for each eclipse/transit which has the steepest portion
of the ramp).

In conclusion, we recommend the 2-D Gaussian for
{\it Spitzer} IRAC Channel 4 sub-array centroiding 
of bright targets as it appears to behave in a near-optimal 
manner.  As we will show, this results in very small scatter in 
the resulting photometry.

\section{Raw photometry} \label{photometry}

\begin{figure}[htb]
\begin{center}
 \includegraphics[width=\hsize]{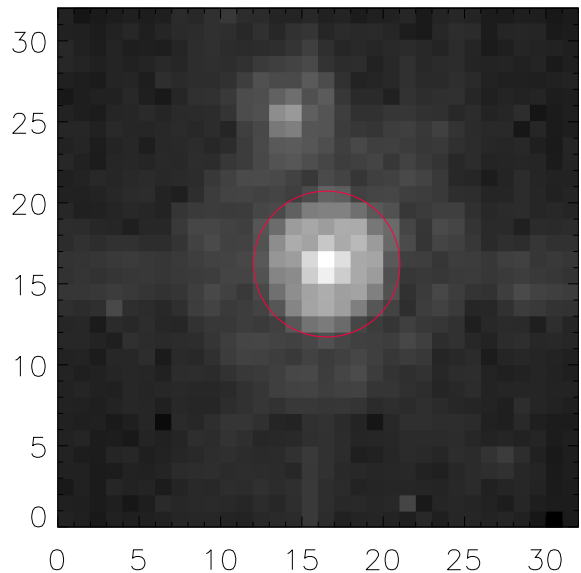}
 \caption{Logarithmic scaling of a medianed image from one of
our observations; horizontal and vertical axes are pixels.  White 
represents about 13,700 counts per pixel, while black is about 
6 counts per pixel.  The red circle is a 4.5 pixel radius aperture.}
\label{snapshot}
\end{center}
\end{figure}

\begin{figure*}[htb]
\begin{center}
 \includegraphics[width=170mm]{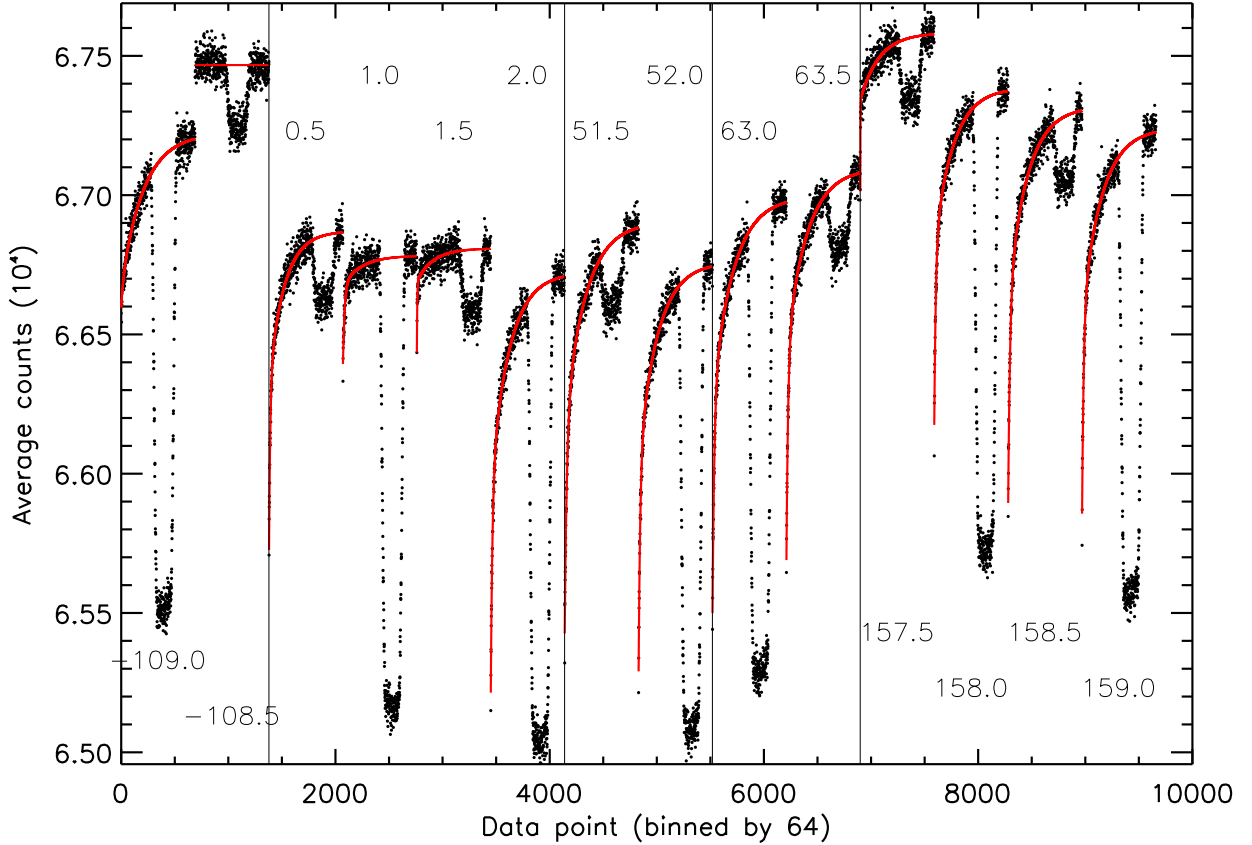} 
 \caption{Atlas of transits and secondary eclipses obtained at 8 microns
with Spitzer.  The photon counts per exposure averaged over 64 exposures
is plotted versus the sequence of each set of 64 exposures.  The numbers 
above or below each transit/eclipse indicate the orbital phase.  The
solid red curves show the best-fit double-exponential ramp model (which
includes the phase-variation of the planet for the secondary eclipses).}
\label{fig02}
\end{center}
\end{figure*}

We carried out photometry on our data using aperture photometry
with a circular aperture.  The contribution of the pixels on the edges 
of the circle are calculated by multiplying the total flux in the
pixel by the geometric fraction of the the pixel that is covered
by the circular aperture. This is done using the GSFC Astronomy 
Library IDL routine {\it pixwt.pro}.  
Figure \ref{snapshot} shows a logarithmically-scaled median image
from one of our sets of observations.  As can be seen in
the image, the sub-array is 32 pixels square, and a companion
M-dwarf lies 9 pixels from the target star. 
We tried a range of apertures, discussed below in \S \ref{aperturesize},
but our final analysis uses a 4.5 pixel radius aperture which
is shown as a red circle centered on the target star.  
Note that this aperture size contains the bulk of the target flux,
and is near the minimum in flux just inside the first
Airy ring; this makes our photometry less sensitive to
variations in position.
We have fit both stars with the measured point response
function for IRAC Channel 4, and we find that the contribution
of the M dwarf within this aperture is less than 0.06\%
of the target star flux for all of our observations.
The resulting light curves for our twelve observations plus
the transit and eclipse from \cite{Knutson2007} are shown
in Figure \ref{fig02} for an aperture of 4.5 pixels radius.
Note that for each transit/eclipse pair, the flux at eclipse
is higher than the flux at transit; this indicates the
planet is brighter on the day side than the night side.

\subsection{Background subtraction} \label{background}

To subtract the background, we used a similar procedure as that described
in \cite{Knutson2007}; namely we fit a Gaussian to a histogram of the 
counts from a subset of pixels located in the corners of the image
(excluding the M dwarf companion, and excluding the top row). This 
background contributes about 1.9-2.6\% of the total flux in our 4.5 pixel 
radius aperture, which we subtract from the time series frame by frame.  As
discussed by \citet{Harrington2007} and in the IRAC Instrument Handbook, 
the flux and the background of the 1st-5th and 58th frame of every set of 64 
exposures is systematically lower than the other exposures in a set.
However, after we carried out background
subtraction frame by frame, the offset in these exposures does not appear in 
our total time series.  Consequently, we believe it is due to a bias offset
that affects the entire frame uniformly, and thus is easily removed.

\subsection{Detector ramp} \label{ramp}

{\it Spitzer} was not envisioned as an instrument for carrying out 
high-precision photometry on bright targets; consequently it
was designed without sub-mmag exoplanet photometry in mind.  An instrumental
artifact that appears at the $\sim$mmag level in photometry, but
can be up to 10\% for low-illumination pixels over 33 hours, is 
the so-called ``detector ramp" \citep{Deming2009}: a pixel which is 
illuminated uniformly in time shows a gradual increase in the detected 
flux (see Figure \ref{fig02}).  This is an important
effect to correct for in fitting photometric time-series;
unfortunately there has not been a full understanding of this
effect for the {\it Spitzer} IRAC detector.  Here we derive a toy-model 
which qualitatively matches the behavior of the ramp (\S \ref{toymodel}).  
We show that prior functions used for ramp-corrections in other analyses 
of IRAC data \citep[e.g.][]{Deming2006, Desert2009} do not have the 
correct functional form to describe the observed ramp (\S \ref{priorramp}).

Understanding how to remove the detector ramp has evolved with time.  
Early work \citep{Deming2006} ratioed the transit star to other sources, 
and modeled the baseline in the ratio as linear or quadratic in time.  
Fitting functions to the ramp directly have used either exponentials in 
time \citep{Harrington2007} or polynominals in the log of time 
\citep{Knutson2009,Desert2009}.  These approaches have been 
adequate at lower signal-to-noise ratios, but for the present high S/N 
data we are motivated to find an improved functional form.
We propose a new functional form, motivated by the toy model, which has
the correct behavior and matches the observed ramps better (\S
\ref{newramp}).  We apply a range of tests to this ramp model, and
show that it performs better than other ramp models (\S \ref{rampperformance}).  

\subsubsection{Toy model for the detector ramp} \label{toymodel}

The ramp effect is hypothesized
to be due to trapping of electrons in detector defects (``charge-trapping").
When a pixel is first illuminated, the charge traps are effectively empty, 
and some fraction of the electrons generated by the incident flux are
retained by the traps instead of being read out by the array.  As these
charge traps fill, the effective gain of the detector goes up, until 
eventually the effect disappears.  Thus
bright, non-variable sources should have a detected flux that asymptotes to 
a constant value.  When a pixel is not illuminated (or illuminated at
very low intensity), the trapped charge gradually releases with time,
causing the charge traps to become empty; this leads to ghost images
after exposure of bright sources.  A consequence of this model is
that higher illumination pixels fill their charge traps more quickly,
thus showing a much shorter detector ramp timescale.  Although it is not 
clear that this model is correct, its predicted behavior agrees 
qualitatively with the observed IRAC photometric properties:  for a 
bright source, the central pixels have a short ramp which saturates 
quickly, while the pixels with lower illumination in the wings of the PSF
show a more gradual ramp.  This behavior, though, is difficult to
model quantitatively as the pixel illumination varies with time
due to {\it Spitzer} pointing variations (see section \ref{centroid}).

A simple toy model can be developed for charge-trapping as follows.  Let
$\gamma_{\rm m}$ be the fraction of volume of a detector pixel filled with
charge-traps, $\gamma(t)$ be the fraction of volume of a pixel with empty 
charge-traps at time $t$, and $\beta$ be the total well-depth  (electrons). 
A pixel is illuminated below saturation with an intensity causing 
$I(t)$ electrons to be released per second, while the measured intensity 
is $I^\prime(t)$ (e$^-$ sec$^{-1}$).
As the pixel is illuminated, the charge traps fill up at a rate proportional 
to the intensity times the fraction of volume of empty charge traps;  
however, there is also a timescale $\tau$ at which electrons 
in charge traps are released, causing $\gamma$ to increase.  
This gives 
\begin{equation}
{d\gamma \over dt} = -{I(t)\over \beta} \gamma
-{\gamma_{\rm m}-\gamma(t)\over  \tau}.
\end{equation}  
The measured intensity is then
\begin{equation}
I^\prime(t) = (1-\gamma) I(t) + \beta{\gamma_{\rm m}-\gamma\over \tau}.
\end{equation}
These equations have no closed-form solution for an arbitrary
$I(t)$; however, we can solve for their behavior in certain limits.
For instance, if the illuminating intensity is constant, $I(t)=I_{\rm 0}$, for
times $t\ge t_{\rm 0}$, then 
\begin{equation}
\gamma(t) = \gamma_{\rm m}(1-I\tau/\beta)^{-1} +
\left(\gamma(t_{\rm 0})-\gamma_{\rm m}(1-I\tau/\beta)\right)e^{(\tau^{-1}-I/\beta)(t-t_{\rm 0})}.
\end{equation}

As we are observing a bright star, we can simplify this equation by
assuming that $\tau^{-1} \ll I_{\rm 0}/\beta$, but we are still below saturation 
and in the linear regime ($I_{\rm 0} t_{\rm exp} \la 0.9 \beta$, where
$t_{\rm exp}$ is the exposure time).  In this limit 
\begin{equation}
I^\prime(t) \approx I_{\rm 0} (1-\gamma) 
= I_{\rm 0} \left(1-\gamma(t_{\rm 0}) e^{-{I_{\rm 0}\over \beta}(t-t_{\rm 0})}\right).
\end{equation}
This gives
the expected ramp behavior: more strongly illuminated pixels have an apparent 
intensity, $I^\prime$, that asymptotes to a constant more quickly on
a timescale $\beta/I_{\rm 0}$.   At modest illumination, this becomes
\begin{equation}
I^\prime(t) = I_{\rm 0} (1-\gamma_{\rm 0} + \gamma_{\rm 0} I_{\rm 0} \beta^{-1} (t-t_{\rm 0}));
\end{equation}
a gradual linear ramp.

In the limit of zero illumination, 
\begin{equation}
\gamma(t) = \gamma_{\rm m}-(\gamma_{\rm m}-\gamma_{\rm 0}) e^{-(t-t_{\rm 0})/\tau}.
\end{equation}
Thus the apparent intensity is 
\begin{equation}
I^\prime(t) = {\beta \over \tau}
(\gamma_{\rm m}-\gamma_{\rm 0})e^{-(t-t_{\rm 0})/\tau}.
\end{equation}
This leads to persistent or ghost images that decay exponentially
with time after observation of a bright target when the illumination 
is so strong that $\gamma_{\rm 0} \ll \gamma_{\rm m}$.

\subsubsection{Prior models for the detector ramp} \label{priorramp}

The correction for the detector ramp is typically applied {\it after} 
performing photometry on the target star rather than at the pixel
level, with some exceptions \citep[e.g.][]{Knutson2007,Laughlin2009}.  There is a 
simple reason for this: at the pixel level it is very difficult to 
disentangle the ramp from pointing variations, while aperture photometry 
with a sufficiently wide aperture gets rid of most of the pointing 
variations and isolates the ramp behavior.  Most ramp corrections have 
simply been functions that match the behavior of the ramp;  two popular
functions are $a_{\rm 0} + a_{\rm 1} (t-t_{\rm 0}) + a_{\rm 2} \log{(t-t_{\rm 0})}$ (log-linear) and
$a_{\rm 0} + a_{\rm 1} \log{(t-t_{\rm 0})} + a_{\rm 2} \left[\log{(t-t_{\rm 0})}\right]^2$ 
(quadratic log) which both seem to work well for IRAC Channel 4 data
\citep[e.g.][]{Deming2006,Deming2009}.  In particular, the logarithmic
term matches the shape of the ramp well, which is steeper towards the 
beginning and shallower towards the end.

Given the toy model described in the prior section, this
logarithmic behavior would appear to be largely coincidental.  
Aperture photometry
combines pixels with a wide range of illuminations; those with
high illumination, which is most of the flux, have a ramp that
is steep and saturates quickly, while those with low illuminations
in the wings of the PSF have a longer timescale ramp that saturates
more slowly.  Summing up these short and long timescale ramps
gives a shape which is steeper in the beginning and more gradual
as time passes, which is well modeled by a logarithmic function.
The linear term or a squared logarithmic term gives enough extra 
degrees of freedom to the model to adjust the slope of the curve 
and give a good fit to most ramp data.  This model has the additional 
advantage that it is linear (except the initial time, $t_{\rm 0}$, in
the logarithmic term), and thus is quick and easy to fit to
the observed ramp.

However, the log-linear and quadratic log ramp models have a serious 
drawback: they do not have the correct behavior on long timescales.
Both the log function and linear function increase without bound,
while the detector ramp does appear to saturate at a constant
value for the brightest pixels.  Thus, with a dataset with
long duration, the log plus linear model or quadratic log ramp
models should do a poor job
in fitting the ramp shape.  In addition, the log plus linear
and quadratic log models do not describe what the final asymptotic flux value
will be, and thus does not give a ramp {\it correction}, but
only gives an empirical fit to how the flux is varying over
the timescale of a given observation.  These points are
particularly important for small aperture photometry where most
pixels have high illumination and thus saturate quickly.
Consequently, we advocate not using ramps that are
polynomials in time and/or log time.

\subsubsection{New model for the detector ramp} \label{newramp}

Motivated by the toy model in section  \ref{toymodel}, we
decided to try an exponential ramp function.  As this model
predicts, the time constant is a function of pixel illumination.
However, due to the pointing variations, we were not successful
in correcting for the ramp on the pixel level.  Instead
we tried a ramp correction function that is simply the sum of two
exponential terms: 
\begin{equation} \label{expramp}
F^\prime/F= a_{\rm 0}-a_{\rm 1} e^{-t/\tau_{\rm 1}}-a_{\rm 2} e^{-t/\tau_{\rm 2}},
\end{equation}
where $F^\prime$ is the flux affected by the ramp, and
$F$ is the flux corrected for the ramp.
Although this does not have exactly the correct behavior for
the sum of pixels with different illuminations (assuming the
toy model is correct), it does have the correct asymptotic
behavior, and qualitatively represents the correction from higher
and lower illumination pixels.  

Figure \ref{fig02} shows the ramp function overplotted with
our data for the fourteen transits and eclipses.

\subsubsection{Performance of double-exponential ramp} \label{rampperformance}

In addition to the qualitatively correct behavior of the double-exponential
ramp, we find that this ramp 
function leads to a smaller scatter in our derived eclipse depths
for the seven eclipses, as well as less sensitivity to the various
choices we make in our analysis.  We held the planet-star
radius ratio and impact parameter fixed at the transit values when
analyzing the secondary eclipses.  We ran initial fits for each ramp
on photometry computed for a 3.5 pixel radius aperture with the first 55 
minutes for each transit/eclipse discarded, and then determined how the 
eclipse depth changed as we varied individual analysis parameters.
The scatter in the seven secondary eclipse depths for the 
double-exponential ramp model is smaller by 30\% than for the 
log-linear ramp (3.05\% versus 3.94\%), and smaller by 20\% compared
to the quadratic log ramp (3.05\% versus 3.68\%).   
An additional indication of the more robust
behavior of this ramp function is that the mean depth only changes
by 0.2\% if we first fit the ramp to the out-of-transit/eclipse
data, and then fit the transit/eclipse to the ramp-corrected data
versus a simultaneous fit to the transit/eclipse and ramp.
The log-linear and quadratic log ramps have a mean eclipse depth
that changes by 0.5\% and 1\%, respectively, between these
two reduction techniques.  Likewise, the double-exponential
ramp changes in eclipse depth by only -1.1\% if the first
55 minutes are discarded, while the log-linear and quadratic
log change by -1.2\% and 3.7\%, respectively. The
double-exponential ramp is also less sensitive to aperture size.
For apertures between 3.5 to 5.0 pixels in radius, the 
individual eclipse depths vary by 1\%, while for the
log-linear and quadratic log ramps, the variation is 1.5\%
and 2.4\%, respectively.
Finally, the total $\chi^2$ for the double-exponential ramp
model is slightly smaller by 21 than the log-linear ramp, and by
27 than the quadratic log ramp, which by the F-test for the additional 
13 free parameters (for seven transits and six eclipses; the phase
-function eclipse has no ramp) favors the double-exponential ramp 
at $>99.999$\% confidence.

There are two drawbacks of the double-exponential ramp function: 
(1) it involves two non-linear fit parameters, $\tau_{\rm 1}$ and $\tau_{\rm 2}$, 
which need to be optimized with a non-linear minimizer; (2)
in some cases when there is very little ramp (possibly due
to high illumination prior to our observations), one or
both of the $\tau$ values can diverge, or in some cases
they can become degenerate.  However, these drawbacks are
straightforward to deal with by setting bounds in a non-linear 
solver, and are outweighed by the improved fit to the observed ramp,
the correct asymptotic behavior, the smaller scatter
in our results,  weaker dependence on aperture size, the 
weaker dependence on whether the ramp is
first corrected or fit simultaneously, and less sensitivity to whether
the steep portion of the ramp is discarded.
Thus, we advocate using this ramp function for IRAC Channel 4 data.

\subsection{Aperture size} \label{aperturesize}

We carried out photometry with apertures ranging in radius from 1 pixel
to 7 pixels.  We fit each transit and eclipse separately, and
then computed the standard deviation of the data divided by the best-fit
model (this is essentially the residuals in magnitudes).  We find
that the residuals are minimized at 4.39 mmag for an aperture radius of 3.5 pixels; this
is the same aperture chosen in \citet{Knutson2007}.  For aperture radii of
3.0 pixels, the scatter is 4.5 mmag, while for 4.0 pixels it is 4.40 mmag,
and for 4.5 pixels is 4.47 mmag, indicating that there is a shallow dependence 
on scatter with aperture size.  

More importantly, we wish to minimize the presence of red noise in
the residuals of the data.  Consequently, we looked at the scatter
in the residuals binned over a range of bin sizes from one exposure
to 1920 exposures as a function of aperture size.  We then took
the mean of the scatter of the binned data over all fourteen transits
and eclipses, and computed the product of this mean scatter divided by
the unbinned scatter over all bin sizes.  The minimum occurs for an aperture 
of 4.5 pixels; although this has a {\it slightly} larger residual scatter 
without binning, the binned residuals are smaller relative to the unbinned
residuals than for the 3.5 pixel radius aperture case.
We have also measured the power spectrum of the residuals, and we
find that the 4.5 pixel radius aperture minimizes the long period power.
Thus, we feel that this aperture size represents an appropriate compromise
between small scatter in the unbinned data (which varies weakly with
aperture size) versus minimization of the red noise component.

\begin{figure}[htb]
\centering
\resizebox{8cm}{!}{\includegraphics[width=\hsize]{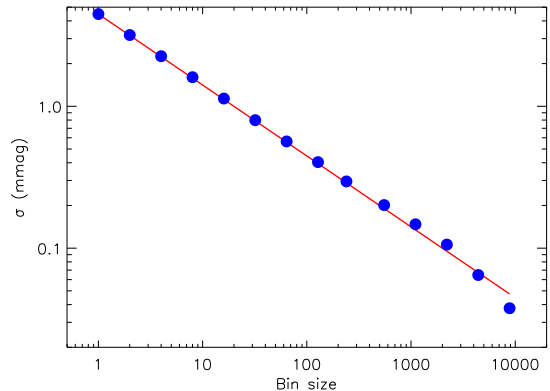}}
\caption[]{
Scatter in the data (vertical axis) divided by the model binned by
a number of bins (horizontal axis), averaged over the 14 transits
and eclipses presented here.  Red line is the extrapolation from
the unbinned data by the inverse square root of the bin size.} \label{sig_vs_bin}
\end{figure}

Figure \ref{sig_vs_bin} shows the scatter in the binned residuals,
averaged over the 14 transits, as a function of bin size.  Even up 
to bin sizes of 8832 exposures (1 hour
bin), the scatter in the data does not deviate significantly from the
inverse square root of the bin-size; this indicates that the residuals
are uncorrelated, and thus there is little (if any) red noise present.
Remarkably the scatter in the one hour bins reaches 30 $\mu$mag; however
this is after subtracting off the double-exponential ramp model.

For an aperture radius of 4.5 pixels, the median counts per exposure
is 66,792.  If photon counting errors dominate, then the expected
noise level is 3.87 mmag per exposure.  Including read noise
(4.5 e$^-$ per pixel) and sky noise ($\sim$30 counts per pixel),
the expected uncertainty is 3.99 mmag (we did not use the BCD
uncertainties since these overpredict the noise properties according
to the Spitzer Observer's Manual).  The standard deviation
in the residuals is 4.47 mmag, which is only 15.5\% greater than
the photon counting error and 12.0\% greater than the expected
errors including read noise and sky noise.  Thus, the noise properties after
correcting for the detector ramp are very close to the expected
photon noise. The 3.5 pixel radius aperture has a residual scatter which
is only 11\% above the photon noise; however this aperture size
appeared to have more significant red noise, so we opted for
the larger 4.5 pixel radius aperture.

\subsection{Conversion to barycentric Julian date} \label{barycentric}

We use the JPL Horizons ephemeris for the {\it Spitzer} orbit to convert the
satellite time (keyword {\sc DATE\_OBS}) to Barycentric Julian
Date (BJD) in Coordinated Universal Time (UTC).\footnote{There is an
additional $+$65.184 sec offset to convert to Barycentric Dynamical Time
for our data.}  This correction is 
important since heliocentric and barycentric Julian Date can differ by as 
much as a few seconds, and different time systems can vary by
seconds depending on the number of leap seconds included, which is close 
to the timing precision we can achieve with these data \citep{Eastman2010}.

\subsection{Error analysis} \label{erroranalysis}

We compute the errors on model parameters by calculating the residuals from
each fit, shifting these by a random number of observations,
and then adding the shifted residuals back to the best-fit 
model, then re-fitting; the so called ``prayer-bead" analysis
\citep[e.g.][]{Agol2007}.
For each transit and eclipse we carried out 2000 iterations
of the prayer-bead shifts.  
This has the advantage of preserving correlations in the
noise of the data that might still be present.  For instance,
if the ramp-model is incorrect, there may be systematic deviation
due to using the wrong ramp model, and these deviations are
preserved within the residuals.  This approach has some
disadvantages; for instance, if the noise behaves differently
within eclipse than outside eclipse, this might exaggerate
the noise outside eclipse.  Another disadvantage of this technique
is that the number of independent trials is limited by
the size of the data set since point-order has to be preserved;
consequently we also randomly chose to reverse the residuals or
change their sign to give more independent noise realizations.
There is also the possibility that the effects of correlated noise
may be removed in the fit.  Even with these disadvantages we
expect that this technique gives a fairly conservative estimate 
for the uncertainties on model parameters.

\section{Fit for stellar and planet variability} \label{stellarvariability}

Stellar variability can affect our fits to the transits and eclipses,
as well as our estimate of the planet variability.   We follow the
approaches in \citet{Knutson2009} and \citet{Sing2009} to derive
a new estimate of the relation between the optical and 8 micron
flux variability of the star by comparing our data set with
that of \citet{Henry2008}, plus additional unpublished data.
The contemporaneous optical monitoring data were taken with the T10 0.8 m 
automated photometric telescope at Fairborn Observatory, which has a median
time sampling of 1 day, but gaps of up to 2 weeks.  The optical time
series consists of the mean of Stromgren $b$ and $y$ magnitudes, 
subtracted from a nearby comparison star, HD 189410, giving  $f_{\rm opt}=
\Delta[(b+y)/2]$ as a function of time, as described in more detail 
in \citep{Winn2007}.  Data outliers are removed (usually taken in 
poor conditions), resulting in a total of 700 observations over 5 seasons.

Using the measured stellar rotation period of $P_{\rm *} = 11.953 \pm 0.009$ days
from \citet{Henry2008},
we fit a sinusoidal function to data within each season to interpolate the
measured optical fluxes to the times of our {\it Spitzer} observations.
For all but one of our IRAC observations there were at least one optical 
observation taken within 1 day, and all within 2 days.
We then computed the total unocculted 8 micron flux, $f_{\rm ir}$, at the mid-transit
and mid-eclipse times of our 14 observations, after correcting for the best-fit 
double-exponential ramp, to look for a correlation between the 
8 micron and optical fluxes.
The initial data seemed to show little correlation between the infrared
and optical fluxes, so we carried out a regression of the infrared
fluxes against five variables:  (1) the optical flux, $f_{\rm opt}$, which
is entirely due to the star; (2) the phase $\Phi$ which determines
whether the source is transiting or eclipsing (i.e.\ whether we are 
seeing the day or night side of the planet), $\Phi = 0$ during transit
and $\Phi=1$ during secondary eclipse; 
(3) the average centroid position, $x_{\rm c}$, on the detector for each
of our observations (the $y$ position varied little
between observations); (4) the average infrared background flux
scaled to our aperture, $f_{\rm bkd}$ (this was already subtracted
in earlier analysis of the data, but we nevertheless include it
in the regression); and (5) the amplitude of the first exponential 
ramp, $a_{\rm 1}$.  The last three terms are included to take into
account the possibility of flat-field errors, imperfect background
subtraction, and the imperfect performance of our ramp function.

\begin{figure}
\centering
\resizebox{8cm}{!}{\includegraphics[width=\hsize]{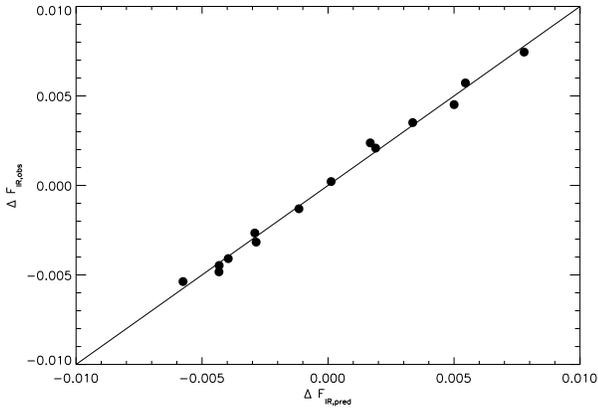}}
\caption[]{Left hand side of equation \ref{ircorrelation} versus
the right hand side.  Solid line is equality; the vertical scatter in the
residuals of this relation is 0.35 mmag.}
\label{regression}
\end{figure}

We find the best-fit relation 
\begin{eqnarray} \label{ircorrelation}
{f_{\rm ir}\over \langle f_{\rm ir}\rangle}-1
&=& (0.197\pm 0.022) {f_{\rm opt}-\langle f_{\rm opt}\rangle \over \langle f_{\rm opt}\rangle} \cr
&+& (1.044\pm 0.026)  {f_{\rm bkd}\over \langle f_{\rm ir}\rangle} 
- (6.59\pm 0.84) \times 10^{-4} x_{\rm c}\cr
&+& (1.19\pm 0.16)\times 10^{-3} \Phi \cr
&+& (1.55\pm 0.48)\times 10^{-6} a_{\rm 1} \cr
&-& (12.7\pm 1.2)\times 10^{-3},
\end{eqnarray}
where $\langle f_{\rm ir}\rangle$ is the ramp-corrected flux averaged
over all 14 observations, while $\langle f_{\rm opt}\rangle$ is the average
over the optical flux at the times of the 14 observations.  The
left hand side of this relation is plotted versus the right hand
side in Figure \ref{regression}; the scatter about this relation is 
0.35 mmag.  We have computed the uncertainties on the regression 
coefficients by Monte Carlo simulation.  

The standard deviation of the residuals of our sinusoidal fits to the
optical data is 2.5 mmag (after exclusion of a few outliers), which
is 1.8 times the optical flux uncertainty  \citep[1.4 mmag;][]{Henry2008}.
Using the sinusoidal fits, the uncertainty on the optical flux we
predict at the mid-points of our observed transits and eclipses ranges
from 0.6-1.4 mmag after inflating the optical errors by a factor of 1.8.
Since infrared stellar fluctuations are 20\% of the optical, this
predicts a scatter of 0.1-0.3 mmag in the infrared, which is consistent
with the measured scatter.

We have computed the
expected spectral change at 8 microns compared to $(b+y)/2$ for
a star spot model in which the star spots are modeled as
Kurucz stellar atmospheres \citep{Kurucz1992} with 4000-4500 K \citep[which
is the estimated temperature from occulted star spots measured
with HST by][]{Pont2008}, while the bulk of the star is 5000 K,
following the procedure described in \citet{Knutson2009}.
We find the expected change at 8 microns is 21-23\% of the change 
in the optical, very close to our measured value, the first
coefficient in equation \ref{ircorrelation}.

There are several important implications of this relation:
(1) as the scatter in this relation is only 0.35 mmag, this
indicates that photometry with {\it Spitzer} is reproducible to 
0.35 mmag over a 590 day period; (2) this scatter limits our uncertainty
on the measurement of the night-side planet flux (during transit),
so we can claim that the night side variability is less
than 0.35 mmag, which is about 17\% of the planet's night-side flux
or 10\% of the planet's day-side flux (which we fit for in section
\ref{eclipsedepth}); 
(3) the transits are 1.19$\pm$0.16 mmag fainter than the eclipses ---after 
accounting for stellar variability--- due to the cooler 
night side of the planet than the day side, confirming the phase 
variation detected in \citet{Knutson2007};  (4) the infrared flux 
variations are about 20\% of the optical variations. 

The derived infrared/optical correlation is nearly twice the value derived
in \citet{Knutson2009}, which used a smaller subset of data
to carry out the correlation and thus was unable to regress
against these other factors.  Our estimate of the
expected flux variations from Kurucz stellar atmospheres indicates
that our derived value is likely correct.
However, \citet{Knutson2009} derived a larger change in the stellar flux 
---by interpolating the observed $y$-band light curve--- than we obtained
by sinusoidal fitting of the $(b+y)/2$ light curve over the period
of duration of the phase function observation, so the resulting
estimates of 8 micron stellar variation for the \citet{Knutson2007}
observation are nearly the same: a 0.6 mmag increase in stellar flux 
between transit and eclipse.

\section{Eclipse and transit models} \label{model}

We fit a model of a straight-lined trajectory
of the planet over the disk of the star.
To compute the transits and eclipses, we used the analytic formulae
from \citet{Mandel2002}, treating the planet as a uniform disk
(no limb-darkening), and the star as a disk with a linear 
limb-darkening law.

For each transit the model has six physical parameters and four 
ramp parameters: the stellar flux $F_{\rm *}$; the sky velocity $v$ 
(units of stellar radius per day); the impact parameter $b$ (units of 
stellar radius); the planet-star radius ratio $p = R_{\rm p}/R_{\rm *}$
(dimensionless); the time of central transit $t_{\rm c}$ (in Barycentric
Julian Day, BJD); the linear limb-darkening parameter of the star
$u_{\rm 1}$ (dimensionless); and the double-exponential ramp parameters
$a_{\rm 1}$, $\tau_{\rm 1}$, $a_{\rm 2}$ and $\tau_{\rm 2}$ (equation \ref{expramp}).
Note that we neglect the contribution of the planet's flux during
transit; this is because we find this is completely degenerate with
the transit depth, and thus leads to problems in fitting
\citep{Kipping2009}.  We initially neglected phase variation of the planet
and variation of the flux of the star during transits as the ramp 
affects all of the transit data sets and thus a short timescale 
(5 hour) planet or stellar variation can not be disentangled 
from the ramp for a single observation.

For the secondary eclipses, we assumed that the planet phase-function 
followed the same {\it shape} as that of \citet{Knutson2007}, which we held 
fixed in our fits to each secondary eclipse, but we allowed the 
{\it total planet flux} to vary for each eclipse.  For each secondary 
eclipse we held fixed the planet/star radius ratio $p$, the impact 
parameter $b$, and the velocity $v$, at the best-fit values from
the transit observations; these parameters are poorly constrained
by the secondary eclipses, and holding them fixed has no
impact on the fitting.  Thus, for each secondary eclipse
we have three physical parameters that
are varied: the stellar flux $F_{\rm *}$; the planet flux $F_{\rm p}$;
the central time of eclipse $t_{\rm c}$; as well as the four ramp parameters.

\subsection{Results from fits to individual transits/eclipses}

\begin{figure}
\centering
\resizebox{8cm}{!}{\includegraphics[width=\hsize]{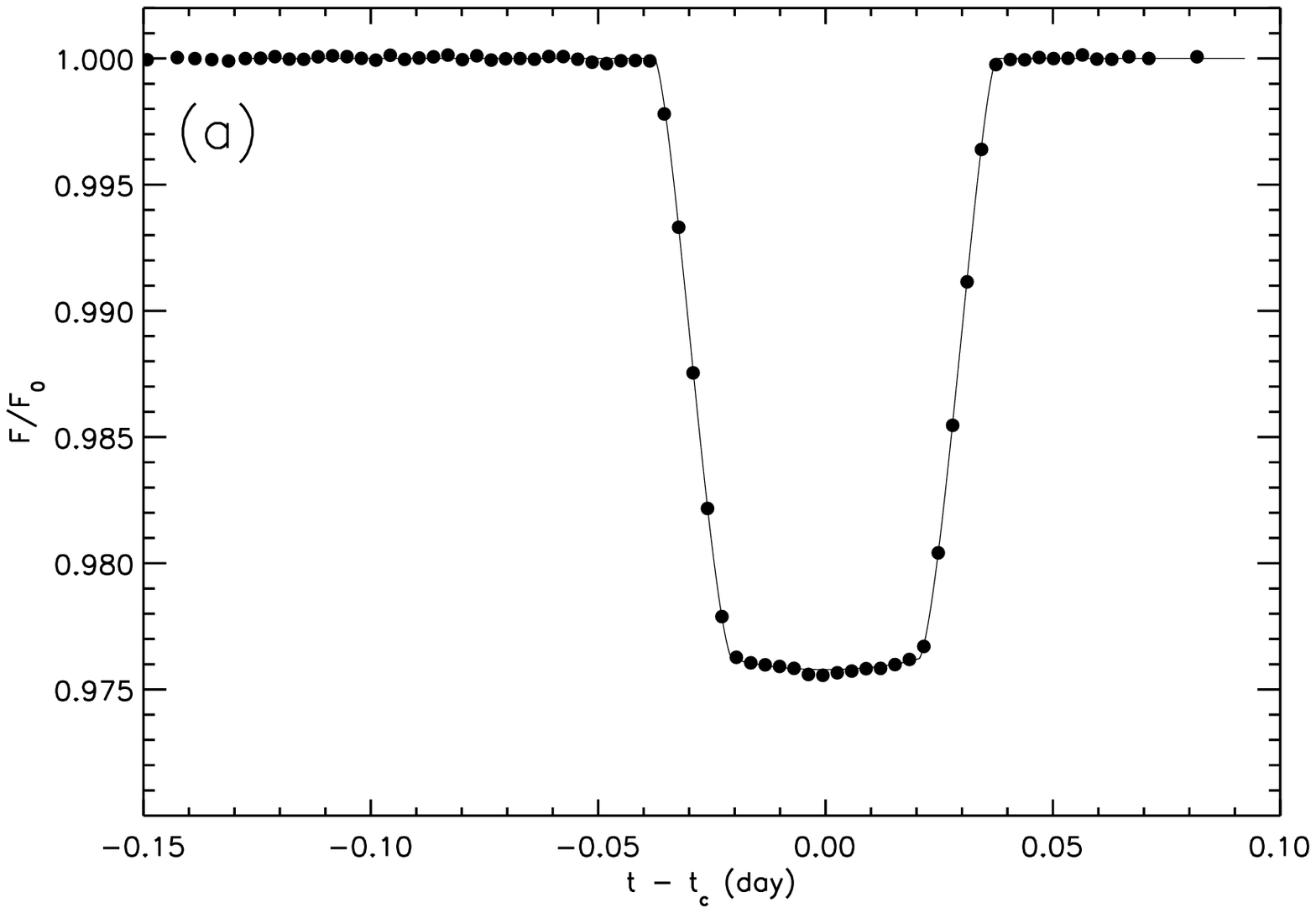}}
\resizebox{8cm}{!}{\includegraphics[width=\hsize]{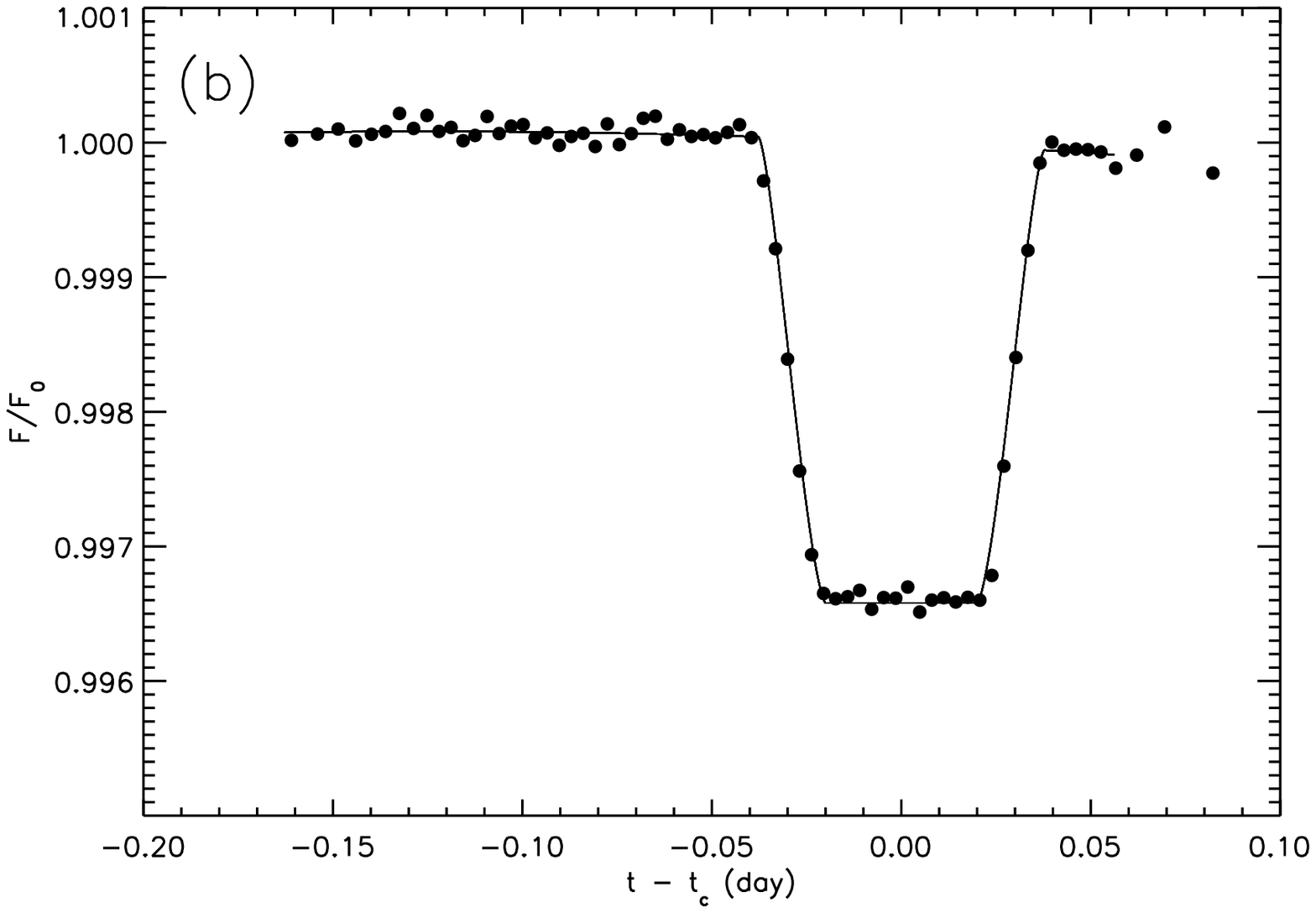}}
\caption[]{(a) Average light curve for the seven transits and (b) seven
eclipses with best-fit models (solid lines).  The data have been
binned by 8960 data points to a total of 69 data points for each.}
\label{average}
\end{figure}

We allowed the model parameters to vary independently for each 
transit/eclipse.  These fits were necessary
since a simultaneous fit to the entire data set is computationally
intensive due to the large number of data points;  we avoided 
pre-binning the data to preserve as much information as possible about 
the noise in the final results.  
Figure \ref{average} shows the average of all seven transits and all 
seven eclipses, corrected for the detector ramp and folded
to the same orbital phase.  We have binned the data by
8960 exposures to 69 data points for clarity.

\subsection{Transit impact parameter and sky velocity}

\begin{figure}
\centering
\resizebox{8cm}{!}{\includegraphics[width=\hsize]{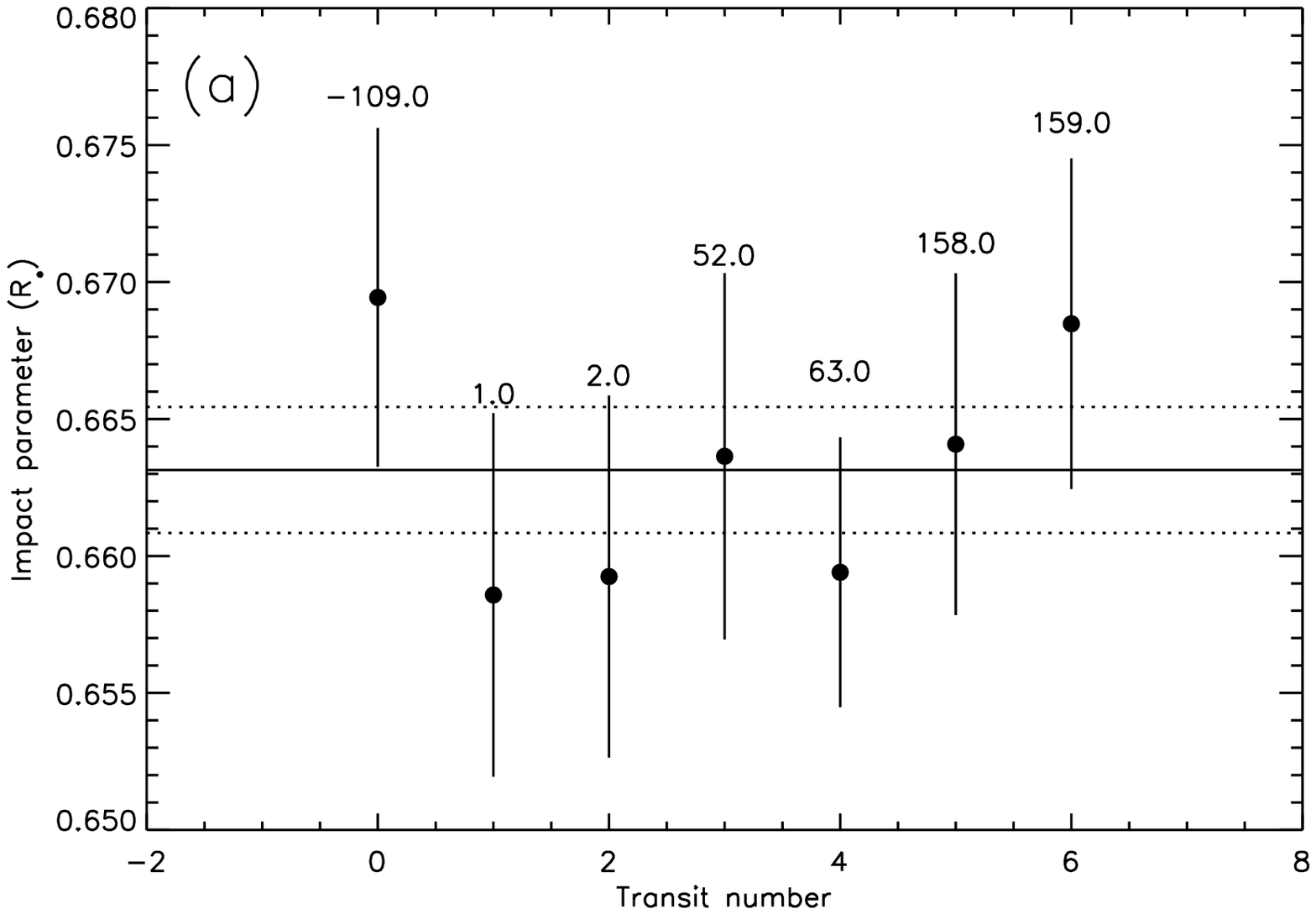}} 
\resizebox{8cm}{!}{\includegraphics[width=\hsize]{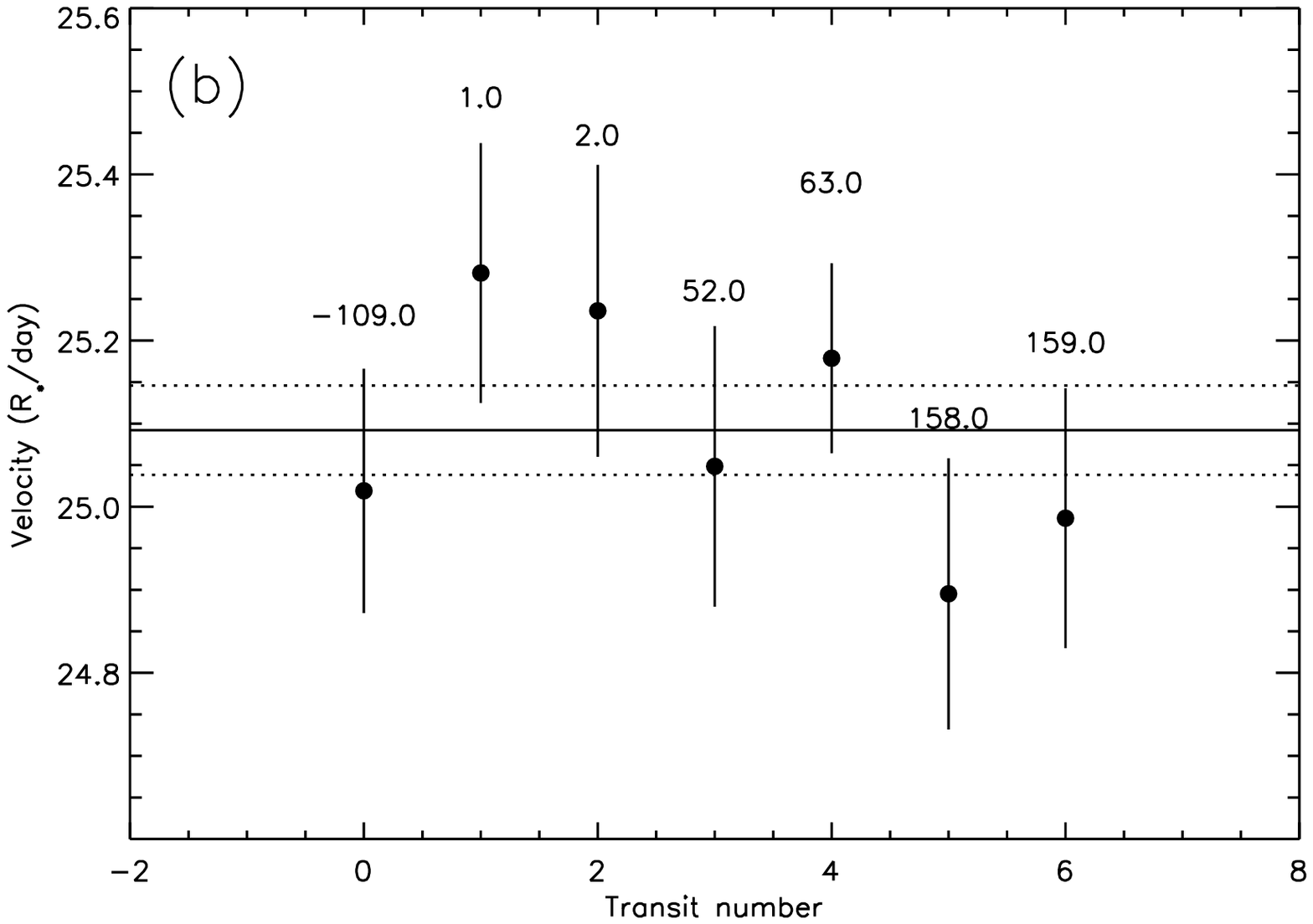}} 
\caption[]{
(a) Impact parameter and (b) transit velocity,  versus
transit number with estimated error bars. The horizontal solid
line is the average of the seven measurements, while the dotted lines
are the uncertainties on the average values.  The numbers above each data point show the
number of periods before or after the zero-point of our
measured ephemeris.}
\label{impact_velocity}
\end{figure}

For the transits, we found that the sky velocity, $v$, and impact
parameter, $b$, have no evidence for variation.  Figure
\ref{impact_velocity} shows each of these parameters plotted versus
the transit number.  The sky velocity has an average 
fractional uncertainty of 0.62\%; the scatter in the measured values 
is 0.57\%, and thus is consistent with being constant.  Combining our 
data together, we find the average sky velocity is 25.125$\pm$0.064
$R_{\rm *}$ day$^{-1}$, while a limit on the variation of the sky
velocity is $dv/dt = (-5.5 \pm 6.6) \times 10^{-4}R_{\rm *}$ day$^{-2}$.
The impact parameter has an average measured value of 0.6631$\pm$0.0023
$R_{\rm *}$, with an average fractional uncertainty for each observation of 0.93\%
and a fractional scatter for the seven observations of 0.67\%; also 
consistent with being constant.  We constrain the 
change in impact parameter to be $db/dt = (-0.02 \pm 2.67) \times 10^{-5} 
R_{\rm *}$ day$^{-1}$. Thus our data indicate that the impact parameter and sky 
velocity of the transits remain constant to $<$1\% over a duration of 590 days.

\begin{table*}[htb] 
\caption{\label{tab01}Transit parameters}
\begin{tabular} {rcccccccccc}
\hline
Phase & $F_*$     & $t_c$        & $\sigma_{t_c}$ & $\Delta t_c$ & $b$     & $v$                 & $(R_p/R_*)^2$& $u_1$ & $-a_1,-a_2$ & $\tau_1,\tau_2$\\
      & (counts)  & (BJD-2454000 days) &   (sec)        &    (sec)     & ($R_*$) & ($R_*$ day$^{-1}$) &  (\%)   &       & (counts)    &    (10$^{-2}$days)      \\
\hline
 -109.0 & 67222.6 &   37.611919$\pm$0.000034 &  3.0 &  -4.2 & 0.6694$\pm$0.0062 &  25.02$\pm$0.15 & 2.4088$\pm$0.0037 &  0.08$\pm$ 0.03 &   75, 550&0.6010,6.6616 \\
    1.0 & 66782.7 &  281.655291$\pm$0.000040 &  3.4 &  -0.0 & 0.6586$\pm$0.0066 &  25.28$\pm$0.16 & 2.4022$\pm$0.0047 &  0.11$\pm$ 0.03 &  281, 107&0.4735,5.4525 \\
    2.0 & 66722.4 &  283.873884$\pm$0.000042 &  3.6 &   1.4 & 0.6592$\pm$0.0066 &  25.24$\pm$0.18 & 2.4253$\pm$0.0063 &  0.11$\pm$ 0.03 &  756, 752&0.4929,5.5313 \\
   52.0 & 66758.8 &  394.802711$\pm$0.000028 &  2.4 &   5.2 & 0.6636$\pm$0.0067 &  25.05$\pm$0.17 & 2.4333$\pm$0.0051 &  0.13$\pm$ 0.03 &  756, 712&0.4891,5.6423 \\
   63.0 & 66995.3 &  419.207003$\pm$0.000036 &  3.1 &   1.7 & 0.6594$\pm$0.0049 &  25.18$\pm$0.11 & 2.4225$\pm$0.0049 &  0.12$\pm$ 0.02 &  773, 722&0.5322,6.2047 \\
  158.0 & 67385.6 &  629.971694$\pm$0.000033 &  2.9 &   1.8 & 0.6641$\pm$0.0062 &  24.90$\pm$0.16 & 2.3984$\pm$0.0062 &  0.16$\pm$ 0.02 &  639, 570&0.4560,5.5904 \\
  159.0 & 67242.2 &  632.190128$\pm$0.000039 &  3.4 & -10.4 & 0.6685$\pm$0.0060 &  24.99$\pm$0.16 & 2.3965$\pm$0.0074 &  0.08$\pm$ 0.03 &  715, 669&0.4890,5.8697 \\
\hline
\end{tabular}
\end{table*}

\begin{table*}[htb] 
\begin{centering}
\caption{\label{tab02}Eclipse parameters}
\begin{tabular} {rccccccc}
\hline
Phase & $F_*$     & $t_c$        & $\sigma_{t_c}$ & $\Delta t_c$    & $F_p/F_*$& $-a_1,-a_2$ & $\tau_1,\tau_2$\\
      & (counts)  & (BJD-2454000 days) &   (sec)        &    (sec)  &  (\%)    &   (counts)  &    (10$^{-2}$days)      \\
\hline
 -108.5 & 67466.9 &   38.722278$\pm$0.000265 & 22.9 &   9.1 & 0.3345$\pm$0.0057 &    0,   0&0.0000,0.0000 \\
    0.5 & 66870.3 &  280.546423$\pm$0.000263 & 22.7 & -32.5 & 0.3469$\pm$0.0060 &  609, 535&0.3793,4.5632 \\
    1.5 & 66808.5 &  282.765713$\pm$0.000350 & 30.2 &  29.3 & 0.3420$\pm$0.0068 &  255, 118&0.1134,4.8741 \\
   51.5 & 66904.4 &  393.693798$\pm$0.000414 & 35.8 & -26.2 & 0.3368$\pm$0.0042 &  750, 727&0.6350,6.1896 \\
   63.5 & 67097.2 &  420.317184$\pm$0.000287 & 24.8 &  16.2 & 0.3623$\pm$0.0064 &  759, 647&0.5089,6.0830 \\
  157.5 & 67582.4 &  628.862649$\pm$0.000390 & 33.7 & -30.8 & 0.3378$\pm$0.0091 &  330, 239&0.0310,5.6520 \\
  158.5 & 67318.3 &  631.081875$\pm$0.000313 & 27.0 &  25.5 & 0.3477$\pm$0.0075 &  737, 685&0.5505,5.6994 \\
\hline
\end{tabular}
\end{centering}
\end{table*}

\subsection{Transit and eclipse times} \label{timingsection}

\begin{figure}
\centering
\resizebox{8cm}{!}{\includegraphics[width=\hsize]{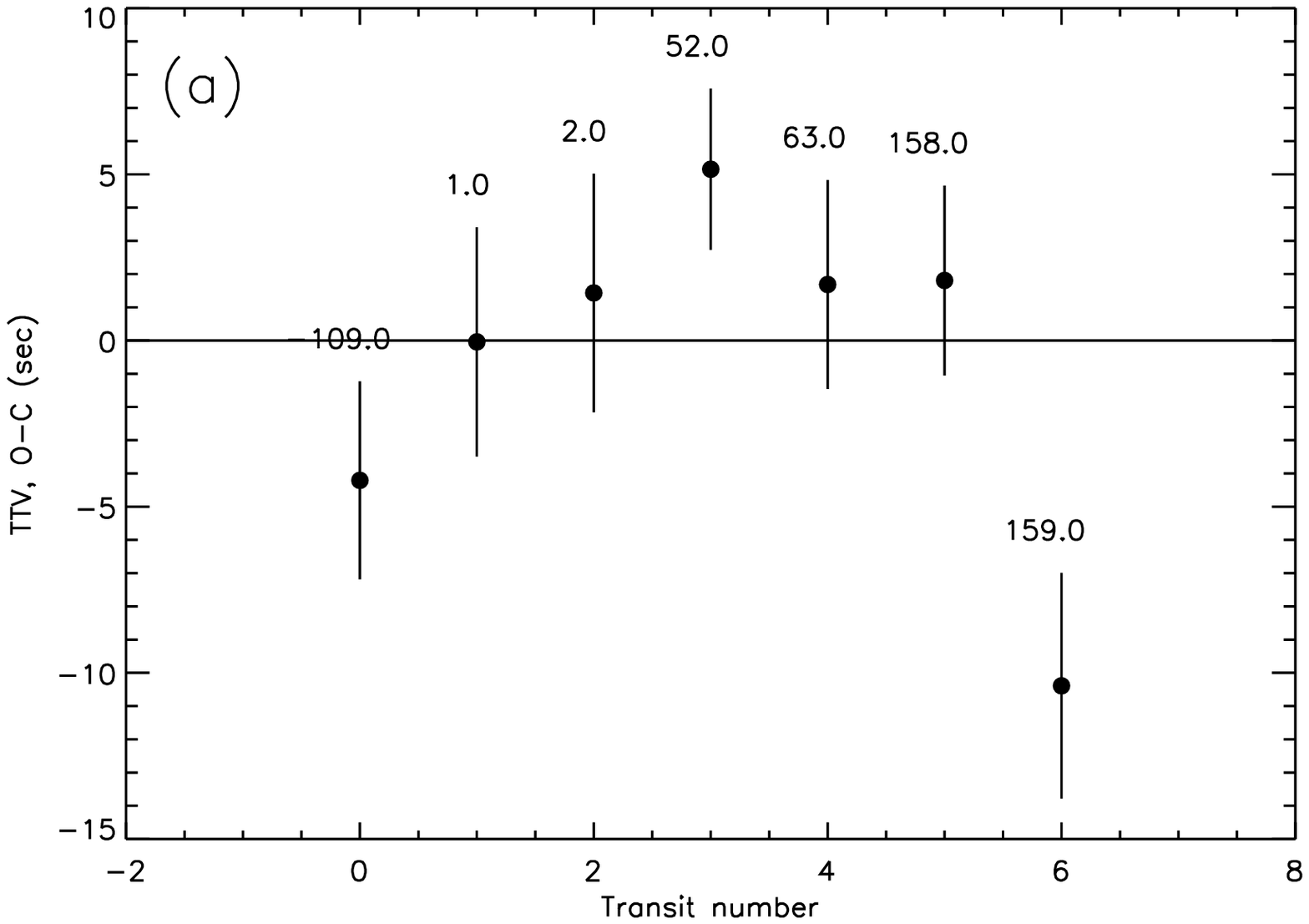}} 
\resizebox{8cm}{!}{\includegraphics[width=\hsize]{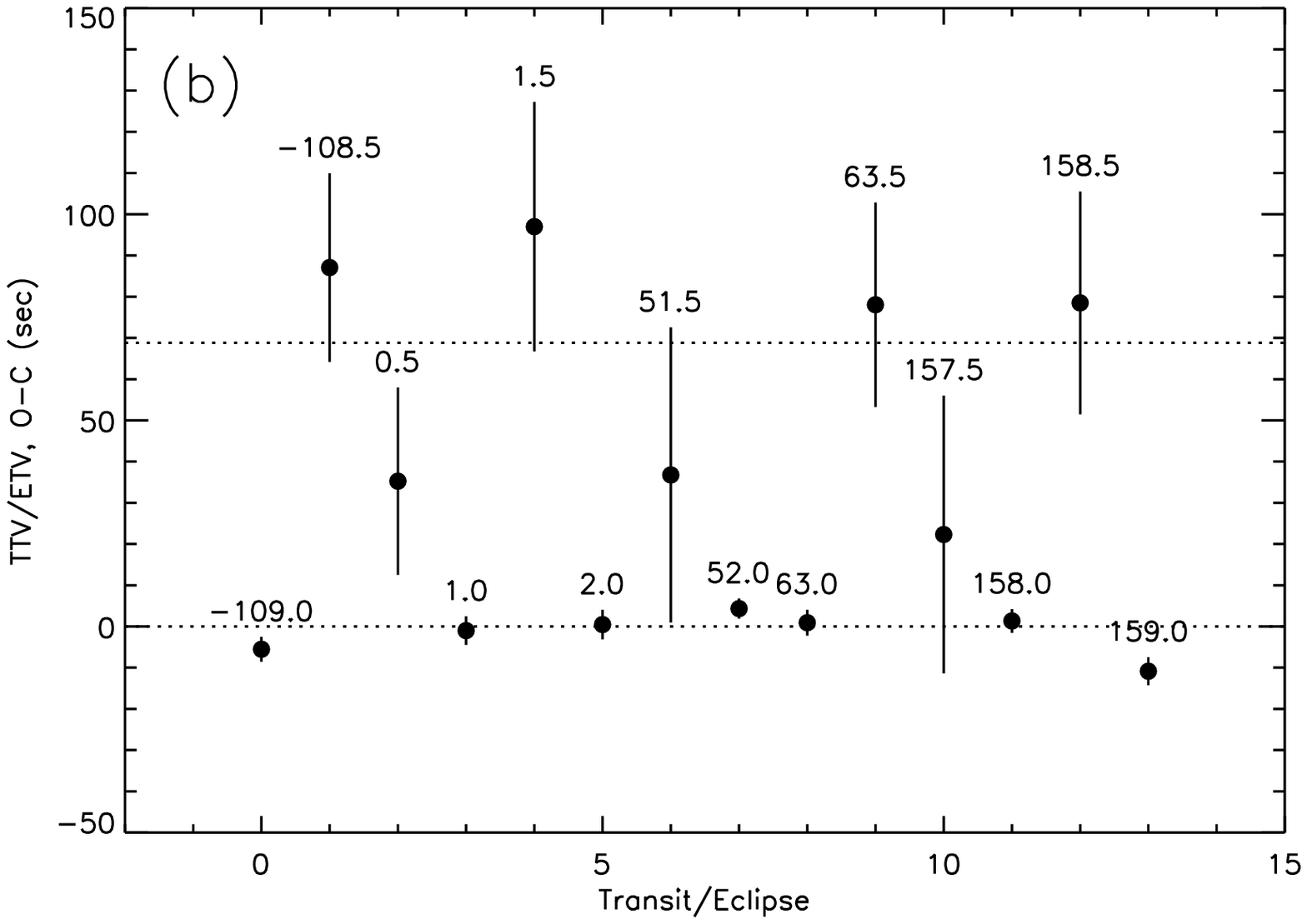}} 
\caption[]{
(a) Transit timing variations, observed minus calculated for a constant
ephemeris, $O-C$, and (b) both transit and eclipse timing
variations (ETV), $O-C$, versus transit/eclipse chronological order with 
estimated error bars.  Note that panel (b) has a vertical scale that 
is $12$ times larger than panel (a).
The horizontal dotted lines are the average of the seven transits and
seven eclipses; this is zero for the transits as we have subtracted off
the best-fit transit ephemeris.  The numbers above each data point show the
number of periods before or after the zero-point of our
measured ephemeris; the points are not plotted as they occur in time,
but are simply evenly spaced.}
\label{ttv}
\end{figure}

We measured the transit and eclipse times for the seven transits and eclipses,
shown in Figure \ref{ttv}, as well as in Tables \ref{tab01},\ref{tab02}.
The errors on the transit times range from 2.4-3.6 seconds and are some of
the most precise transit times ever measured; comparable to, or better than, the 
three {\it HST} transit times reported in \citet{Pont2007}.  We fit
separate ephemerides to the transits and eclipses; the results are shown
in Table \ref{tab03}.  If we instead fit the transit times with
the quadratic function: $t_{\rm n} = t_{\rm 0} + P n + \frac{1}{2} \dot P P n^2$, where
$t_{\rm n}$ is the time of the $n$th transit, and $\dot P$ is the change
in period of the orbit, we find $\dot P = -0.06 \pm 0.02$ sec yr$^{-1}$.
Since this is primarily due to the last data point, which may be an outlier,
we do not view this as a significant detection.

The uncertainty on the transit times and eclipse
times, as well as the derived ephemerides, are inversely proportional to depth
of the transits and eclipses.  This is due to the fact that the timing precision
is proportional to the inverse of the flux gradient with time during ingress and
egress.  The ingress and egress duration are the same for the transit and 
eclipse (assuming a circular orbit), so the ratio of the flux gradient scales 
with the ratio of their depths.  The ratio of the depths is proportional 
to the ratio of the surface brightness of the star to the surface brightness 
of the planet (limb-darkening is weak for
this star at 8 microns), so the transit time precisions are smaller than
the eclipse precisions by the ratio of the surface brightness of the planet
to the star, which is about 14.3\%, or a factor of $7.0$.

Figure \ref{ttv} shows the deviations from the ephemeris for both the
transits and the eclipses. The transits have a scatter of 5.1 seconds,
which is very close the the observational errors; there is no evidence
in our data for transit timing variations over a period of 590 days.
The eclipses also appear to be precisely periodic - their scatter with
respect to the best-fit ephemeris is 27 seconds, which is comparable
to the errors on each data point.  The period derived from the transits 
differs from that derived from the eclipses by only 0.1 seconds!

The eclipses appear 69$\pm$11 seconds later than 1/2 of an orbital period
after the transits.  As discussed in \citet{Knutson2007}, this is
partly due to the light-travel across the system, 30.8$\pm$0.6 seconds
\citep[this uncertainty is due to the uncertainty in stellar mass,
$M_{\rm *} = 0.806 \pm 0.048 M_\odot$,][]{Torres2008}, while the remaining 
38$\pm$11 seconds can be mostly accounted for by the hot spot on the planet causing
an offset in the time of eclipse when the planet is modeled as a uniform
disk, as shown in section \ref{timingoffset}.

\begin{table}[htb] 
\caption{\label{tab03}Transit and eclipse ephemerides}
\begin{tabular} {lcc}
\hline
& $T_0$   & $P$ \\
& (${\rm BJD_{UTC}}$) & (days) \\
\hline
Transit: & 2454279.436714$\pm$0.000015 & 2.21857567$\pm$0.00000015 \\
Eclipse: & 2454279.437510$\pm$0.000125 & 2.21857456$\pm$0.00000131 \\
\hline
\end{tabular}
\end{table}

\subsection{Limits on the presence of companion planets from transit-timing}
\label{ttvsection}

These transit data show no significant timing variations, but from these we 
can constrain the maximum mass allowed of additional planets in the system.  Transit timings are 
a particularly sensitive probe for planets in or near mean-motion resonance (MMR) 
and previous studies have ruled out Earth mass or super-Earth mass planets in 
low-order MMR for several systems.  Prior transit timing variation (TTV) analyses 
of the HD189733 system 
\citep{Hrudkova2010,MillerRicci2008} used data with timing precision of order 30 
seconds and were sensitive to planetary masses of near (and below) one Earth mass 
in favorable MMRs.  Our {\it Spitzer} observations of HD189733 have nearly a 
factor of 10 better timing precision and consequently have improved sensitivity 
to secondary planets by that same factor.  Here we calculate the maximum mass 
that an additional planet could have based upon these transit data.  To do so 
we note that the $\chi^2$ per degree of freedom of the timing residuals is 
slightly more than unity.  We therefore scale the timing uncertainty by a factor
of 1.5 and then multiply by two to achieve our 2$\sigma$ ($\simeq$ 95\% 
confidence level) upper bound on the timing variations.

\begin{figure}[htb]
\centering
{\includegraphics[width=84mm]{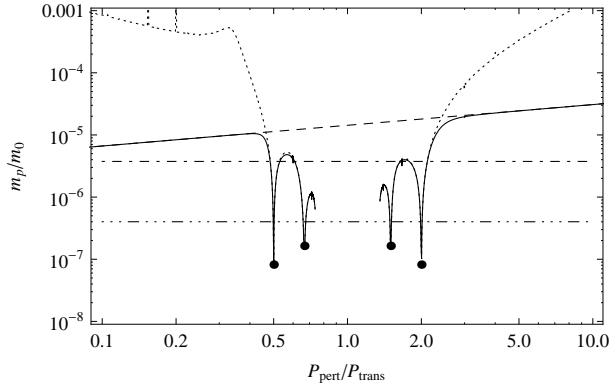}}
\caption[]{Constraints (95\% confidence level) on initially circular orbiting secondary 
planets in HD189733 as a function of the period ratio of the known planet based 
upon these transit data and the radial velocity measurements presented in Boisse 
2009.  The dotted curve are the limits from a TTV analysis alone from equations 
(A7) and (A8) in \citet{Agol2005}.  The dashed line is the expected sensitivity from 
33 RV measurements with 3.5 m/s precision calculated using equation (2) from 
Steffen \& Agol (2005).  The solid curve is the overall sensitivity from both RV 
and TTV measurements (summed in quadrature); the region above this curve is
excluded.  The solid dots are the variation in mean-motion resonance,
$\approx P_{\rm trans}m_{\rm p}/m_{\rm t}$, where $m_{\rm t,p}$ are the masses of the transiting and perturbing 
planets \citep{Agol2005}.
Finally, the horizontal dot-dashed and triple-dot-dashed lines
correspond to the mass of the Earth and the mass of Mars, respectively.} 
\label{ttvlimits}
\end{figure}

Figure \ref{ttvlimits} shows 95\% confidence-level constraints on secondary planets 
with near circular orbits in this system based upon these data and the radial velocity 
(RV) measurements from Boisse 2009.  These limits are derived from the analytic 
formula given in \citet{Agol2005} and \citet{Steffen2005}.  We do not attempt 
an in-depth numerical analyis of these transit times here---the robustness of limits 
derived from analytic formulae was demonstrated in \citet{Agol2007}, \citet{Nesvorny2009}
and \citet{Nesvorny2010}.  These data exclude planets above 2 Earth masses for any orbit 
that lies closer to the known planet than either the interior or exterior 2:1 MMR.  
The transit-timing mass exclusion is superior than the exclusion from radial-velocity
data for periods from 1 to 5 days, excluding all planets with masses greater
than 3 Earth masses within this range.  In addition they exclude planets with masses 
well below the mass of Mars---approximately 0.2 Mars masses or 2 Moon masses at 95\%
percent confidence---in circularly orbiting 2:1 or 3:2 MMRs (interior or exterior).  
For non-circular orbits the sensitivity generally increases. However, in low-order MMR 
the mass sensitivity can decline as much as a factor of 10 for eccentric orbits --- 
\citep[see, for example,][]{Agol2007}.  

\subsection{Effect of hot spot on secondary eclipse time} \label{timingoffset}

As discussed in \citet{Knutson2007}, the 8 micron phase function
indicates that the hottest point on the planet is offset from the
sub-stellar point.  This was predicted by \citet{Cooper2005},
and is attributed to the advection of energy by a super-rotating wind
encircling the equator of the planet.  This offset hot spot means
that the ingress and egress of the secondary eclipse will have
a shape that differs from our model, which utilizes a uniform
disk.  In particular, this means that the steepest portion of
ingress and egress will be offset from the uniform disk case;
since the hotspot is on the trailing side of the planet with respect
to the direction of motion, this causes a delay in the eclipse 
time when fit with a uniform disk model \citep{Charbonneau2005,Williams2006}.  
In \citet{Knutson2007} we estimated that the hot spot would cause a delay 
of at most 20 seconds; however, the fit to the phase function in that
paper did not correct for stellar variation which caused the
location of the hot spot to be underestimated, leading to
an underestimated uniform time offset.

To estimate the magnitude of this effect, we used
a simplified model of the longitudinal planet brightness
which is discussed in \citet{Cowan2010b}.
Briefly, each position on the planet is treated as a parcel
of gas which moves eastward, absorbing star light as it passes across the
day side, all the while radiating with a time constant $\tau_{\rm rad}$.
This model can be parameterized by a single parameter,
$\epsilon= \tau_{\rm rad}/\tau_{\rm adv}$, where $\tau_{\rm adv}$ is
the time it takes a parcel of gas to circle the planet.
Small values of $\epsilon$ (``instant" re-radiation) lead to 
darker night sides and day side temperatures which are
in equilibrium with the incident stellar flux.  Large
values of $\epsilon$ lead to nearly uniform temperatures
at each latitude.  Thus in the small or large $\epsilon$
limits we expect no timing offset since the day sides
are symmetric.  Only with $\epsilon \sim 1$ is there
an offset hot spot causing a phase function which
peaks before secondary eclipse, as well as a slight
offset in the times of eclipse ingress and egress
if fit with a uniform planet.

\begin{figure}[htb]
\centering
{\includegraphics[width=84mm]{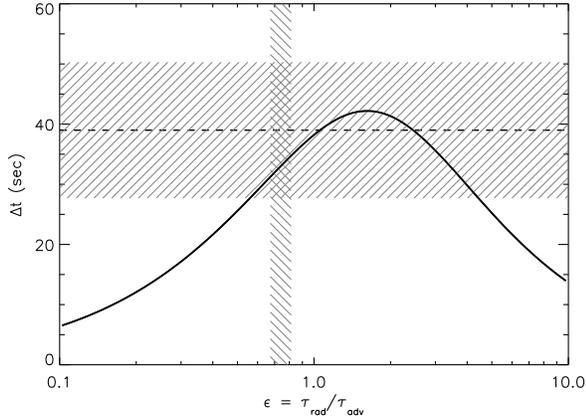}}
\caption[]{Timing offset for a hot spot model as a function
of the ratio of the radiative to advective time scales. The dashed
line is the best-fit eclipse time offset after correction for
light travel time, and horizontal rectangular shaded region is the 1-$\sigma$
confidence limit on this time.  The vertical rectangular shaded region
is the best-fit value of $\epsilon$ to the 8 micron phase
function, after correcting for stellar variability.} \label{hotspotoffset}
\end{figure}

We computed the effect of $\epsilon$ on the time
of secondary eclipse by solving for the planet day-side longitudinal
surface brightness at the equator in the Rayleigh-Jeans limit 
and assuming a constant temperature with latitude.
We computed the eclipse ingress and egress from
this model for the planet surface brightness, we fit
this simulated eclipse light curve with a model for the eclipse of a 
uniform planet, and from this best fit we determined the 
offset in the time of eclipse, the so-called ``uniform
time offset" defined by \citet{Williams2006}.  Figure
\ref{hotspotoffset} shows this time offset as
a function $\epsilon$; a positive offset means that
the secondary eclipse occurs later than expected for
a uniform planet.  The maximum offset predicted by 
this model is 43 seconds, which agrees with the observed 
eclipse time offset.  Our measured eclipse time
is plotted as a dashed line in this figure, with
the uncertainty indicated by the horizontal shaded
rectangle.

\begin{figure}[htb]
\centering
{\includegraphics[width=84mm]{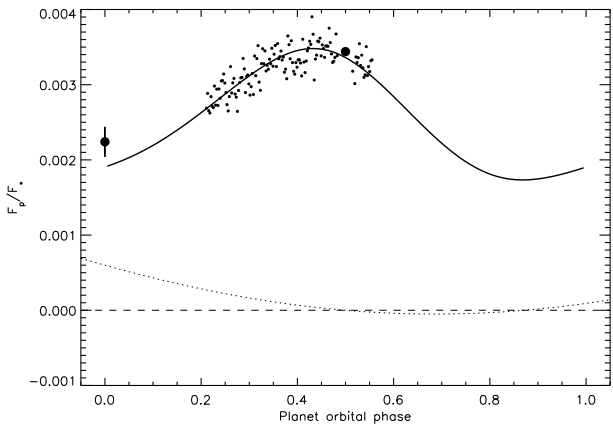}}
\caption[]{Planet phase function after correction for stellar
variability versus planet orbital phase.  We only use the
last $\sim$ 2/3 of the phase function to avoid problems
with the ramp correction, and we masked the secondary eclipse.
The thick solid line is the best-fit model for planet
variability with $\epsilon=0.74$. The dot with error bar on
the left is our estimate of the night-side brightness from
equation \ref{ircorrelation}. The dotted line shows our correction
for stellar variability during the phase function.} \label{phasefunction}
\end{figure}

\begin{figure}
\centering
{\includegraphics[width=\hsize]{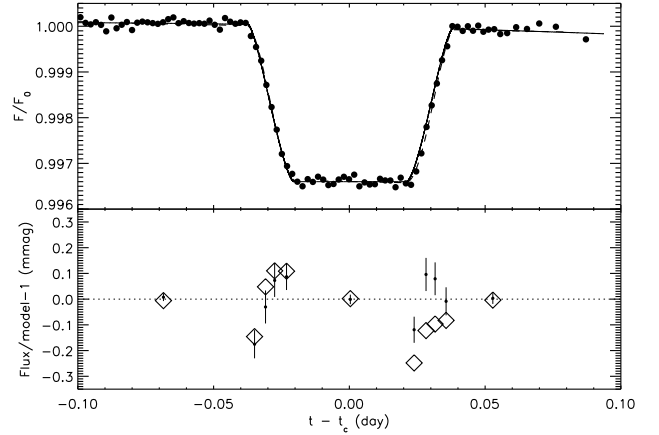}}
\caption[]{Plot of average of 7 eclipses (top panel)
with best-fit uniform planet model, offset by 30.8 seconds
after 1/2 orbital period after the transit ephemeris
(solid line), as well as the $\epsilon
=0.74$ model with light-travel time delay (dashed
line).  The bottom panel shows the residuals binned
into eleven bins: pre-ingress, post-egress, eclipse, 
and four bins each in ingress and egress;  the residuals 
are plotted for the uniform planet model (diamonds)
and $\epsilon=0.74$ model (filled circles with error
bars).}\label{eclipseoffset}
\end{figure}

The location of the peak in the planet phase function 
from \citet{Knutson2007} provides another constraint on 
the location of this hot spot, or equivalently on the
value of $\epsilon$ (see Figure \ref{phasefunction}).  
We used the relation between the infrared and optical 
stellar variability derived in section \ref{stellarvariability}
to derive the change in stellar flux at 8 micron
during the phase function measurement, about 0.6 mmag
over 26.6 hours.  We then
fit the last 2/3 of the measured 8 micron phase function
to estimate $\epsilon$ (Figure \ref{phasefunction}); 
we discarded the first 1/3 of the phase function data 
since it is strongly affected by the ramp correction.
We find a best-fit value of $\epsilon = 0.74 \pm 0.07$,
which we have also plotted as a vertical shaded
region in Figure \ref{hotspotoffset}. This value
of $\epsilon$ predicts a timing offset of 33 seconds,
which is consistent with the measured offset of
$38 \pm 11$ seconds.  Figure \ref{eclipseoffset}
shows a direct comparison of the secondary eclipse
to the average of our seven secondary eclipses.
The top panel shows the binned data as well as
the best-fit secondary eclipse at 1/2 orbital
period after transit plus the 30.8 second light travel
time delay (solid line), as well as the $\epsilon
=0.74$ model with light-travel time delay (dashed
line).  The bottom panel shows the residuals binned
into eleven bins: pre-ingress, post-egress, eclipse,
and four bins each in ingress and egress;  the residuals
are plotted for the uniform planet model (diamonds)
and $\epsilon=0.74$ model (filled circles with error
bars).  The uniform planet model shows points which
are on average higher in ingress and lower in egress,
which is a sign of the shifted hot spot.  The
hot spot model provides a better fit to the data, although
there is still scatter in the residuals which just
reflects the low significance of the eclipse hot spot
detection (the uniform time offset is only 3.5$\sigma$:
38$\pm$ 11 sec).  Note that we have not optimized the hot 
spot model, but only computed the light curve from the best
fit to the phase function (which gives $\epsilon=0.74$).

Consequently, there is no evidence for non-zero 
$e \cos{\omega}$ in this system.  For the estimated
value of $\epsilon$, the remaining time offset
is $6 \pm 11$ seconds, which yields $e \cos{\omega} 
= 0.00005 \pm 0.00009$.

Another prediction of this model is the
day-night brightness difference.  For $\epsilon = 0.74$,
we predict a night-side brightness which is 57\% of the day-side
brightness (day defined at mid-eclipse; night at mid-transit).
This corresponds to a decrease in brightness from the
day to night side which is 0.15\% of the stellar flux, which is
very close to the value of 1.2$\pm$0.2 mmag derived
in section \ref{stellarvariability}.  The minimum
planet brightness (for the visible hemisphere) divided by the maximum planet
brightness for this best-fit model is 50\%,
while the peak in observed planet brightness is
23 degrees, 0.065 orbital periods, or 3.5 hours before 
the secondary eclipse.  On the planet, the hottest longitude is
13 degrees from the sub-stellar point; note that this
differs from the hemispherically integrated peak due
to asymmetry in the longitudinal day-side intensity.  Figure
\ref{phasefunction} shows the phase function data
after correction for stellar variation and binned by 1000 
data points, versus planet orbital phase with the 
best-fit model for the planet variability, $\epsilon=0.74$,
overplotted.  Also plotted is our estimate of the night
side flux based on equation \ref{ircorrelation}.

\subsection{Transit depth variation} \label{transitdepth}

\begin{figure}[htb]
\centering
{\includegraphics[width=84mm]{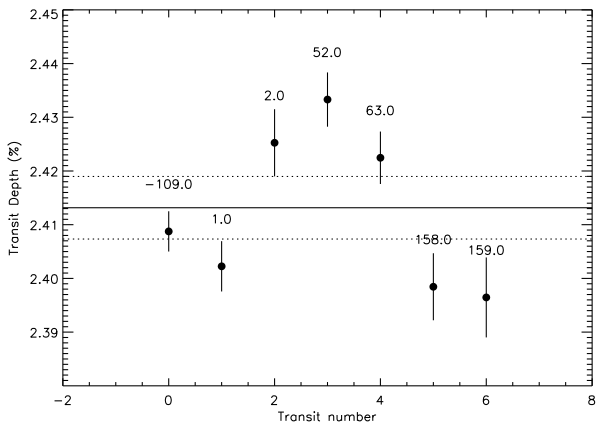}}
\caption[]{Transit depths measured for seven transits.  Horizontal
solid line measures the weighted mean transit depth, while dotted lines
give 1-$\sigma$ uncertainty on this average value based on the scatter
in the data.  The numbers above each point indicate the orbital 
ephemeris.} \label{transitdepths}
\end{figure}

In the fits to the transits we allowed the depth of each transit
to vary independently through the ratio of the planet to stellar
radius, $p=R_{\rm p}/R_{\rm *}$.  This ratio also affects the duration of ingress
and egress, but the amount of time spent in ingress or egress
is small, so the primary effect is on the depth of transit.
Figure \ref{transitdepths} shows the derived
transit depths, $p^2$, for the seven observed transits.   There is
evidence for variability in the transit depth - the uncertainty on the
individual transit depths ranges from 37-74 $\mu$mag, while the
scatter is 145 $\mu$mag.  This corresponds to a scatter in
the fractional variation of transit depth of 0.6\%, while the
ratio of maximum to minimum is 1.5\%.   Fitting the transit 
depths with a single value of $p$ gives a $\Delta \chi^2$ of 40.8 for 6 degrees 
of freedom.  Thus, transit depth variation is detected with high significance.

We allowed the limb-darkening parameter to vary for each transit,
which ranged from 0.08-0.13 for the seven transits, with an
overall mean of $u_{\rm 1} = 0.12\pm 0.01$.  Even though limb-darkening
is weaker in the infrared, an LTE  Kurucz model
atmosphere \citep{Kurucz1992} with parameters close to the values 
inferred for HD 189733, $T_{\rm eff} = 5000$ K, $[Fe/H] = 0.0$, and 
$log[g\rm{(cm/s^2)}] = 4.5$, predicts a linear limb-darkening
coefficient of 0.136, close to what we measure.
We checked that the variation in transit depth is
not due to variations in the best-fit limb-darkening parameter by holding the 
limb-darkening coefficient fixed at the mean value;  this did not affect
our measured values of transit depth.

In fact, the variation in transit depth is not necessarily due
to a change in $p$.  The average depth of each transit is given
by $\delta = \langle I_{\rm path}\rangle \pi R_{\rm p}^2/ (F_{\rm *} + F_{\rm p})$
\citep{Mandel2002}, where $\langle I_{\rm path}\rangle$ is the average 
surface brightness within the path of the planet across the star
and $F_{\rm p}, F_{\rm *}$ are the planet and stellar flux during transit.
Although our model assumes a linear limb-darkening law, if
the path of the planet passes over a region of the star with
brighter than average surface brightness, then a larger depth
will be inferred.  

So, it is possible that the change in transit depth is due
to changes in $\langle I_{\rm path}\rangle/F_{\rm *}$, $R_{\rm p}$, $R_{\rm *}$, or
$F_{\rm p}/F_{\rm *}$.  Variations in $R_{\rm p}$ or $R_{\rm *}$ seem unlikely to be
responsible for the transit depth variation as this would
require changes in radius of 0.3\% for either body: either 220 km
for the planet ($\sim$ 1/3 of a thermal scale height) or
1600 km for the star.  Both of these variations seem too large
to be physically plausible.  Variations in $\langle I_{\rm path}\rangle/F_{\rm *}$ 
due to fluctuations in $F_{\rm *}$ are ruled out as the transit depth 
variations are uncorrelated with the optical stellar flux variations.  
Variations in $F_{\rm p}$ (the night-side planet flux) are less than 
0.35 mmag relative to $F_{\rm *}$, which is too small by a factor of 17 
to be responsible for the transit depth variations.  Thus, the most 
likely possibility is that the transit depth variations are due 
to a variation in the occulted stellar intensity, $\langle I_{\rm path}\rangle$.
This requires only a variation of 0.6\% in the surface
brightness of the path of the planet relative to the average
stellar surface brightness, which is much smaller than
the 12\% change in surface brightness from center to limb
inferred for the best-fit limb-darkening.  We have checked that
individual star spots are not responsible for this variation
by computing the standard deviation of the residuals
in transit divided by the square root of the counts, which 
is 1.153, versus the same quantity computed out of transit,
which is 1.154, so there is no evidence for individual
star spots causing this difference.
Similar inferences have been drawn for Channels 1 and 3
by \citet{Beaulieu2008}, while star spots are easier to detect 
in the optical where the contrast between star spots and the 
stellar disk is larger, and have already been detected for this 
star by \citet{Pont2007}, who also resolved the shape
and color-dependence of the spots.

\subsection{Eclipse depth variation} \label{eclipsedepth}

\begin{figure}[htb]
\centering
{\includegraphics[width=84mm]{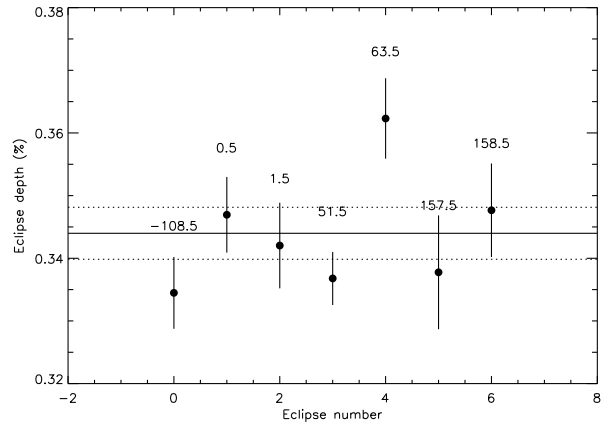}}
\caption[]{Eclipse depths measured for seven planet eclipses.  Horizontal
solid line measures the average transit depth, while dotted lines
give 1-$\sigma$ uncertainty on this average value.  The numbers
above each point indicate the orbital ephemeris.} \label{eclipsedepthfig}
\end{figure}

Figure \ref{eclipsedepthfig} shows the measured eclipse depth
in units of the stellar flux at mid-eclipse.  The weighted mean 
eclipse depth is 0.344 $\pm$ 0.0036\% and the $\chi^2$ fit to the eclipse
depths with this mean value is $14.6$ for 6 degrees of freedom.
The errors on each eclipse depth vary between 0.004-0.009\%, while
the scatter in the depths is 0.01\%, so there is no detection of 
significant eclipse depth variability.   This scatter corresponds
to 2.7\% variation of the mid-eclipse planet brightness; this
can be taken as an upper limit on the planet variability
at 68\% confidence.

\begin{figure}[htb]
\centering
{\includegraphics[width=84mm]{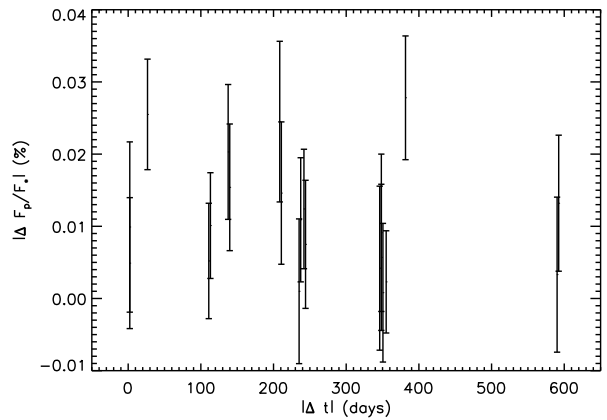}}
\caption[]{For each pair of eclipse observations, we show here the change 
in eclipse depth as a percent, versus the time between the observations.  
If the eclipse depths showed time-correlation ---due to either astrophysical 
variability on some characteristic timescale or detector systematics--- this 
plot would show a rise.  The flat distribution is consistent with Gaussian 
variations at the level of a few percent.} \label{eclipse_variability}
\end{figure}

Although our eclipse depths do not exhibit significant variability, 
they only probe variability on timescales shorter than the baseline 
of the observations (roughly two years). To verify that our data do 
not show any time structure, we plot in Figure \ref{eclipse_variability} 
the change in eclipse depth against the time between observations, for 
each pair of observations ($N\times (N-1)/2 = 21$ for $N=7$). We add uncertainties 
in quadrature to estimate the uncertainty on the flux differences.  
The resulting locus is flat, showing that measured eclipse depth is 
uncorrelated with the time of the observation.  Note that this sets a 
limit not only astrophysical variability scenarios, but also systematic 
errors.

This limit on the variation in eclipse depths is sufficient to 
rule out the predicted variation computed for HD 189733b by 
\citet{Rauscher2008}.  The most extreme prediction they make
is for their $\eta=0.05$, ${\bar U} = $ 800 m s$^{-1}$ model
which has a standard deviation of $\sim$ 8\% in the day-side 
brightness, with the largest excursions of 20\%.   The
largest difference in brightness we see is 8\%, while
the scatter in planet brightness is 2.7\%, so by both
measures the observed variation is a factor of $\sim 3$
smaller than the predictions of this particular model.

Other models predict smaller variations in the day-side
brightness, such as \citet{Showman2009a} who compute
the 8 micron brightness variation for HD 189733b should
be less than 1\%.  Our upper limits are consistent with
this model, but unfortunately do not constrain the
model due to our uncertainties that are larger than
the predicted variations.

With upper limits on both day and night-side variability, it is worth asking 
which of these puts stronger constraints on the planet's physical properties.  
Consider the simple model of \citet{Cowan2010a}, which parametrizes the planet's 
day and night-side brightnesses in terms of the planet's Bond albedo and 
recirculation efficiency.  One can treat brightness variability 
---of the day or night--- as being due to changes in albedo and/or changes in 
recirculation efficiency, and compute how these affect the day and night
side brightness.  We find that the day side variability upper limit
provides a better constraint on variation in the Bond albedo or recirulation 
efficiency than does the night side variability limit, which is $\sim 6\times$ 
larger than the day-side limit.

\begin{figure}[tb]
\includegraphics[width=84mm]{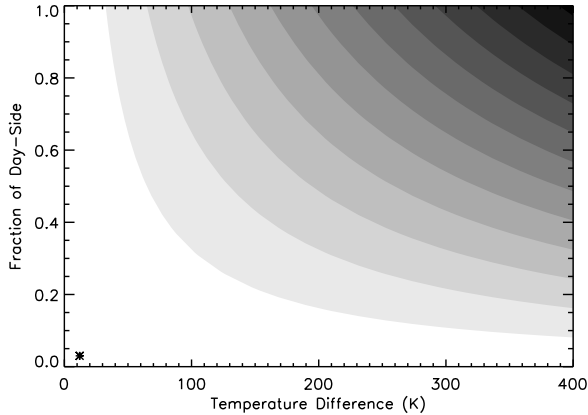}
\caption{An exclusion plot for the covering fraction and temperature contrast of a 
putative storm on the day-side of HD~189733b. The top right corner of the plot is 
excluded at $5\sigma$. For comparison, Jupiter's Great Red Spot has a filling fraction 
of roughly 0.03, and a temperature 12 K cooler than the rest of the planet (asterisk).}
\label{storm_exclusion}
\end{figure}

Another possible origin of planet day-side variability are transient local
variations in the surface brightness, for example due to large-scale ``storms."
We use a toy model where the planet's day side has a uniform
temperature, $T_{\rm d}$, except for a storm with covering fraction $0<f<1$ (the y-axis) 
and temperature difference $\Delta T$ from mean day-side temperature (the x-axis).  
Figure~\ref{storm_exclusion} shows exclusion limits on the largest putative storm 
that could form or dissipate without appearing in our data.  
The increasingly dark shades of gray denote areas of parameter space excluded at 
$1\sigma$ through $5\sigma$.  According to \citet{Showman2009b}, the radius of 
deformation for HD 189733b is $0.3$ of the planetary radius, an order of magnitude larger 
than for Jupiter.  The covering fraction for a typical storm on such a planet would be 
$f=0.1$, for which we cannot rule out storms differing by 324~K (68\% confidence) from 
the average day-side temperature.  For comparison, Jupiter's Great Red Spot has a filling 
fraction of roughly 0.03, and a temperature 12 K cooler than the rest of the planet.  
The bottom line is that our data rule out only the most extreme weather fluctuations on 
HD~189733b.

\subsection{System parameters}

Due to the high precision of our data and weak limb darkening
in the infrared, we can considerably improve the determination
of certain stellar and planet parameters for this system
from our data.  Since both the small inferred value of 
$e\cos{\omega}$ and theoretical predictions indicate that $e$ 
should be close to zero, we set $e=0$ in deriving the system 
parameters.  The uncertainties on the stellar and planet 
parameters are computed for each transit or eclipse
by computing the system parameters from the model parameters 
from each simulation \citep[using the relations in][]{Winn2010}, 
computing the standard deviation of the results from the 
simulations as an estimate of the errors on each parameter, 
and then taking a weighted mean of all transits/eclipses to 
obtain the final mean value of the best-fit parameters.

\begin{table}[htb]
\caption{\label{tab04}Best fit system parameters}
\centerline{
\begin{tabular}{llc}
\hline
Parameter & Best fit & units \\
\hline
$a/R_*$ & 8.863$\pm$0.020 & \\
$b/R_*$ & 0.6631 $\pm$ 0.0023 & \\
$i $    & 85.710 $\pm$  0.024 & deg \\
$e \cos{\omega}$&  0.000050$\pm$ 0.000094& \\
$ u_1$ & 0.118 $\pm$ 0.010 & \\
$\rho_*$  & 2.670 $\pm$ 0.017 & g cm$^{-3}$ \\
$ R_p/R_*$ & 0.155313 $\pm$ 0.000188 & \\
$F_p/F_*$ & 0.3440 $\pm$ 0.0036 & \% \\
$g_p$     &  2145.9 $\pm$    13.5& cm s$^{-2}$ \\
$\rho_p$  & 0.943 $\pm$ 0.024 & g cm$^{-3}$ \\
\hline
\end{tabular}
}
\end{table}

Table \ref{tab04} presents the system parameters determined
from all 14 transits and eclipses.  We have focused on
parameters that are most directly constrained from the
photometry, which are either dimensionless, or have units
of density.  Compared to the values derived in \citet{Torres2008} 
and \citet{Pont2007}, the uncertainties on our values are
smaller by a factor of 2-10.  For the planet surface
gravity, $g_{\rm p}$, we use the velocity semi-amplitude 
$K= 200.56 \pm 0.88$ m s$^{-1}$, derived by \citet{Boisse2009}, 
and for the planet density, $\rho_{\rm p}$, we use the stellar mass
$M_{\rm *} = 0.806 \pm 0.048 M_\odot$ given in \citet{Torres2008}, 
propagating the uncertainties assuming they are uncorrelated 
and Gaussian.

\section{Discussion and Conclusions} \label{conclusions}

The analysis of fourteen transits and eclipses in this
paper has made several improvements to the data reduction
and modeling; in particular, we have found a better function
for fitting the detector ramp of IRAC Channel 4, a
double-exponential.   The
scatter in the residuals is approaching that of photon
counting errors, similar to the precision achieved in other 
IRAC observations \citep[e.g.][]{Todorov2010}, but for a brighter
source star, and the residuals show very little evidence of red 
noise.  These technical developments
have allowed us to make a better correction for stellar
variability, and have given us better constraints on the
parameters of this system.

As HD 189733 is the first planet system in which the phase
variation has been measured at high significance, it provides
some of the tightest constraints on the atmospheric dynamics
of an extrasolar planet.  Our observations of an additional
six transits and eclipses presented here allows us to place 
additional constraints on the longitudinal brightness
distribution of the planet at 8 microns.  In particular, we 
have improved the measurement of the correlation 
between the optical, $(b+y)/2$ band, and 8 micron variations 
in the star over the correlation measured in \citet{Knutson2009},
giving a correlation that agrees better with predictions of star 
spot models.  This measured correlation allows us to derive
a better correction for the stellar variation during the
observation of \citet{Knutson2007}, giving
us a better measurement of the planet's phase function.
In particular, we find that the peak planet flux at 8
microns occurs 3.5 hours before secondary eclipse, which is
1.2 hours before the value derived in \citet{Knutson2007}
without correction for stellar variation; this is consistent
within the errors given in that paper, which were dominated
by the ramp correction.   This measured
phase function predicts a 33 second delay of the secondary
eclipse when fit with a uniform planet model, which is
consistent with the 38$\pm$ 11 second delay that has been
measured after correcting for light travel time across the
system.  This confirms that the phase variation is indeed
due to the planet, and gives a crude eclipse mapping of the planet
detected the 3.5$\sigma$ level,
as first pointed out by \citet{Williams2006}.  It is significant 
that ---for the same high quality photometry--- phase function 
mapping \citep{Cowan2008} is more effective at locating the planet's 
primary hot spot than eclipse mapping \citep{Williams2006}. This is 
because the duration of eclipse ingress or egress is shorter by a 
factor $\sim R_{\rm p}/(2\pi a)$ than the planet's orbital period, 
while the changes in brightness used by both techniques are comparable.  
The superior leverage of phase function mapping will become even more 
marked as interest shifts towards smaller planets in longer orbits.  
That said, the two mapping techniques suffer from different degeneracies
and different impacts of systematic errors and stellar variability, 
so when possible one will want to use both.  We also confirm 
the phase variation by measuring the difference between the 
fluxes at transit and eclipse, and we find the night side is 
fainter by 1.2$\pm$0.2 mmag, or about
64\% of the brightness of the day side.  All of these
constraints are consistent with a model
in which the gas circulating the planet has a radiative 
cooling timescale which is comparable to the advection timescale;  
we find $\tau_{\rm rad}/\tau_{\rm adv} \sim 0.74$ by fitting the
phase function.

The larger offset in the time of peak planet flux
is also in better agreement with the predictions of \citet{Showman2009a}
who found that to obtain agreement with the smaller offset 
of \citet{Knutson2007} they required an inner boundary of 
their atmosphere that was rotating more slowly than synchronous 
rotation;  instead of a sub-synchronous core, this may be
indicative of slower wind speeds due to magnetic drag near
the 8 micron photosphere \citep{Perna2010}.  The sub-synchronously 
rotating and 5$\times$ solar abundance models of \citet{Showman2009a} 
predict a peak brightness at 8 microns which
is 20-30 degrees before secondary eclipse, which agrees well
with our new estimate of 27 degrees.  The same models
also predicts a day-side brightness (mid-transit) which is
0.33-0.35\% of the star's brightness, consistent with our measured
value of 0.344$\pm$0.004\%.  The night side brightness at 8 
micron predicted by the models is 0.17-0.18\% of the stellar brightness, 
which is consistent with our measured value of 0.22$\pm$0.05\%.
Their models also predict very small variations in the secondary
eclipse depth of less than 1\%, which is consistent with our
upper limit of 2.7\%.  The lack of variation of the atmosphere
indicates that the assumptions used in creating longitudinal
maps of planets from phase functions are likely valid \citep{Cowan2008}.

The time delay for the secondary eclipse can be completely accounted for
by the light travel time of the system and delay of ingress
and egress due to a hotspot on the planet which is
offset longitudinally.  Consequently there is no
evidence for a non-zero $e\cos{\omega}$, and we can place
a limit of $e \cos{\omega} = 0.00005 \pm 0.00009$.
If the orbit of this planet is nearly circular, which the
small value of $e \cos{\omega}$ would indicate, then the
interior is likely also synchronously rotating, which seems 
to agree with the \citet{Showman2009a} predictions for the 
phase function.

We have detected 8 $\mu$m limb-darkening of the star at high
significance, $\sim 10\sigma$, which agrees with predictions
of stellar atmosphere models.  However, the individual 
transits vary in depth, which we hypothesize may be due
to variation of the stellar surface brightness that
is occulted by the planet.  This is not surprising given
the strong optical variations of this star which indicate
a significant presence of star spots.  This variation
needs to be accounted for in creating spectral absorption
profiles of transiting planets.  If the data taken are non-
simultaneous, the variation in stellar surface brightness
could affect the inferred depth of transit differently
at different wavelengths, leading to systematic errors
in comparison to model predictions; even simultaneous
data might be affected by the star spot color.  This is a 
stronger effect at shorter wavelengths;  for example, the
contrast in surface brightness of 4000 K star spots
in the IRAC Channel 1 (3.6 $\mu$m) and Channel 2 (4.5 $\mu$m) 
should be 20-40\% higher than for Channel 4 for this star;
thus fluctuations in transit depth could approach 2\% in
these bands.  In addition, this limits the possibility of 
constraining the variations in transit depth due to planet 
oblateness \citep{Carter2010}.

Due to our highly precise transit times spaced over a wide range
in time, the ephemeris we derive is one of the most precise
for any transiting planet.  The high precision is due to the weak 
limb-darkening, stable instrument (thanks to the Earth-trailing 
orbit of Spitzer which leads to stable thermal properties and no 
occultation of targets by the Earth, as occurs with the Hubble 
Space Telescope), allowing a 3-second precision for transit times.
Our ephemeris has a precision
that is $> 10$ times better for the period than that reported in
in \citet{Pont2007}, and agrees with their reported period
within $\sim$2.8$\sigma$: their period is longer by 
0.46$\pm$0.17 sec.  Our ephemeris predicts times of transit
that are -3.3$\pm$5.0, 3.5$\pm$5.0, and 12.6$\pm$3.5 seconds after 
their three transit times in \citet{Pont2007}.  We detect no strong
evidence for transit timing variations in our data, and we estimated 
from analytic formulae the upper mass limits on the presence of
companion planets in this system, improving upon the limits
placed by \citet{MillerRicci2008} and \citet{Hrudkova2010} by a
factor of $\sim 10$.  Theories of the evolution of short-period
planets due to tidal effects and interaction with turbulence
in the protoplanetary disk indicate that they should evolve out 
of mean-motion resonance, so the lack of detected transit-timing 
variations may not be surprising, especially for interior perturbing
planets \citep{Fabrycky2009,Adams2008,Terquem2007,Papaloizou2010}.

In sum, the excellent stability of the {\it Spitzer Space
Telescope}, and in particular Channel 4 of the IRAC camera,
has enabled near photon-limited photometric errors, and
sub-mmag variations over a period 1.6 years.  This has
enabled a better calibration of the contribution of
stellar variability at 8 micron, which has allowed us
to measure the night-side planet brightness, and has
shifted our estimate of the peak of the phase function.

\acknowledgements

This work is based in part on observations made with the Spitzer 
Space Telescope, which is operated by the Jet Propulsion Laboratory, 
California Institute of Technology under a contract with NASA. 
Support for this work was provided by NASA through an award issued 
by JPL/Caltech. E.A.\ acknowledges the hospitality of the Institute 
for Theory and Computation at the Harvard-Smithsonian Center for 
Astrophysics, the Michigan Center for Theoretical Physics, and the 
Kavli Institute for Theoretical Physics where portions of this 
work were completed. H.A.K.\ is supported by a fellowship from
the Miller Institute for Basic Research in Science. This research 
was supported in part by the National Science Foundation under Grant 
No.\ NSF PHY05-51164 and CAREER Grant No.\ 0645416.

\bibliography{bibliography_agol2010}

\end{document}